\documentstyle[prl,twocolumn,aps]{revtex}

\begin{document}
 
\tolerance 50000


\title{The Matrix Product Approach to Quantum Spin Ladders} 
\author{  J.M. Rom{\'a}n$^{1}$,
G. Sierra$^{2}$, J. Dukelsky$^{3}$,  
M.A. Mart{\'\i}n-Delgado$^{4}$,
} 
\address{ 
$^{1}$Departament d'E.C.M., Universitat de Barcelona, Barcelona, Spain.\\
$^{2}$Instituto de Matem{\'a}ticas y 
F{\'\i}sica Fundamental, C.S.I.C.,
Madrid, Spain.\\
$^{3}$Instituto de Estructura 
de la Materia, C.S.I.C.,Madrid, Spain.
\\ 
$^{4}$Departamento de
F{\'\i}sica Te{\'o}rica I, 
Universidad Complutense. Madrid, Spain.
}

\maketitle 
\widetext

\vspace*{-1.0truecm}

\begin{abstract} 
\begin{center}
 \parbox{14cm}{}
\end{center}
\end{abstract}

\pacs{
 \hspace{2.5cm} 
PACS number: 75.10.Jm}

\narrowtext

\section*{Abstract}

We present a manifestly rotational invariant formulation
of the matrix product method valid for spin chains 
and ladders. We apply it to 2 legged spin ladders 
with spins 1/2, 1 and 3/2 and different magnetic 
structures labelled by the exchange coupling
constants, which can be ferromagnetic or antiferromagnetic
along the legs and the rungs of the ladder
We compute ground state energy densities, correlation lengths and
string order parameters. We present numerical evidence 
of the duality properties of the 3 different non ferromagnetic
spin 1/2 ladders. We show that the long range topological order
characteristic of isolated spin 1 chains is broken by the interchain 
coupling. 
The string order correlation function
decays exponentially with a finite correlation length that we compute.
A physical picture of the spin 1 ladder is given in terms of 
a collection of resonating spin 1 chains. Finally for ladders
with spin equal or greater than 3/2 we define a class of AKLT 
states whose matrix product coefficients are given by 9-j symbols.

\section*{I) Introduction}

The matrix product method (MPM) 
is a variational approach appropiate
to study the ground state and excitations of a variety of
one-dimensional lattice systems in Condensed Matter and 
Statistical Mechanics. The 
theoretical and experimental
interest of these systems
has grown spectacularly in the last years, due to the
discovery of interesting and
unexpected physical properties in spin chains and ladders.

The basic idea
behind the MPM is the 
construction of the ground state and excitations
of 1D or quasi 1D systems in a 
recursive way, by relating the states of the system
with length $N$ to that of length $N-1$. 
This simple idea has appeared 
in the past in different places. 
First of all, in the Wilson's
real space renormalization group the 1D-lattice  
is built up by the  addition of a single  site at every RG step \cite{Wil}. 
This procedure is also used
in the density matrix renormalization
group  method (DMRG) of White \cite{Whi}.
Other source of the MPM is 
the well known AKLT state of the spin  1 chain \cite{AKLT}. This is a simple 
but non trivial example of a matrix product state, which has 
motivated various generalizations as the ones of Klumper et al
\cite{K}, Ostlund and Rommer \cite{OR}, etc. 
We shall follow in this paper the formulation of the MPM
due to the latter authors, 
which is based on 
the analysis of 
the fixed point structure of the DMRG ground 
state in the thermodynamic limit \cite{OR}. A closely related
approach  is that of Fannes et al \cite{Fan}. 
The MPM offers an alternative
formulation 
of the DMRG method in the regime 
where the latter  reaches a fixed
point after many RG iterations \cite{D}.

Whereas the DMRG is a purely
numerical method,  the MPM offers the possibility of an analytical 
approach to elucidate the actual structure of the 
ground state (GS) and excitations. 
The MPM is a standard variational method which determines
the variational parameters by minimizing the GS energy. 
Minimization problems are in general  harder
than diagonalization ones. In this 
respect the MPM is so far less performant than the DMRG. 
However we believe that 
the analytical insights gained with the MPM could be used 
to boost the numerical precision and applications of 
both the MPM and the DMRG.

In this paper we apply the MPM to the 2-leg spin ladder.
Spin ladders with diagonal couplings have been studied
with the MPM of Klumper et al in ref. \cite{B}.  
Spin ladders were first studied 
as theoretical labs to test ideas
concerning the crossover from 1D to 2D, with the surprising result
that this crossover
is far from being smooth: the even and odd ladders display
quite different properties converging only 
when the number of legs goes to infinity (for
an introduction to  the subject see \cite{DR}). Even spin ladders 
are spin liquids with a finite spin gap and finite
spin correlation length, while odd spin ladders belong
to the same universality class than the 
spin 1/2 antiferromagnetic Heisenberg chain,
which has no gap and the correlations decay algebraically. 
Another reason to study ladder systems is that 
materials actually exists 
with that structure and hence the
theoretical predictions can in principle be compared with
experimental data concerning the spectrum, susceptibility, etc.

We study in this paper five different spin ladders characterized
by their local spin  $S= 1/2, 1 $ and 3/2 and 
the signs of the exchange of coupling constants
along the legs $J_{\parallel}$ and the rungs $J_{\perp}$. 

In the case of the spin 1/2 ladders we discuss the following
topics: i) the RVB picture of the antiferromagnetic ladder,
ii) the  equivalence between the ladder state and the Haldane 
state of the  spin 1 chain, and
iii) the  duality properties relating the different magnetic 
structures.

In the case of the spin 1 ladder we show that the
long range topological  order characteristic of 
isolated spin 1 chains disappears and the string
correlator decays exponentially with a finite
correlation length.

The study  of the spin 3/2 ladder motivates
the definition of an AKLT state 
characterized  in terms of 9-j symbols.

We compute
GS energy densities, spin correlation lengths 
and string order parameters
and compare our results with those existing in the 
current literature.

The organization of the paper is as follows. In section
II we review the MPM. 
In section III
we particularize the MPM to systems which are
rotational invariant, where the use of  
group theory leads to a  simplification
of the formalism.   
In section IV we present our 
numerical results concerning five different spin ladders.
In section V we summarize our results 
and present some prospects of our work. 
There are 3 appendices which contain 
technical details or proofs of  results 
presented in the main body of the paper.

\section*{II) Review of the Matrix Product Method}

Some  of the results presented in  this
section  are known and can be found in references 
\cite{OR,Fan}. We also present a full account 
of the formulas and derivations 
used in reference \cite{D} where the MPM was 
compared with the DMRG method in the case of the antiferromagnetic
spin 1 chain.

Let us consider a spin chain or a ladder ${ B_N}$ with
open boundary conditions, where $N$ 
denotes the number of sites of a chain or 
the number of rungs  of
a ladder.
To describe the low energy 
properties of ${ B_N}$
one introduces a collection of $m$ states 
$\{|\alpha\rangle_N\}_{\alpha = 1}^{m}$,
which form an orthonormal basis, i.e.
$_N \langle \alpha| \alpha'\rangle_N = \delta_{\alpha, \alpha'}$. 
In the DMRG these states are the most probable
ones to contribute to the GS of the superblock
${B_{N-1}} \bullet \bullet  
B^{R}_{N-1}$ of length $2 N$, formed  by 
adding two sites (or rungs) $\bullet  \bullet$,
and a mirror image $B^{R}_{N-1}$
to the original
lattice ${B_{N-1}}$. 
The basic assumption of the MPM is that the basis
associated with $B_N$ and $B_{N-1}$
are related in a simple manner 
by the  equation

\begin{equation}
|\alpha \rangle_{N} = \sum_{\beta, s} A_{\alpha, \beta}[s] \;
|s\rangle_{N} \otimes
|\beta \rangle_{N-1} \;\;\; , N \geq 2  
\label{1}
\end{equation}

\noindent 
where $|s \rangle_N \; ( s= 1, \dots, m^*)$
denotes a complete set of  $m^*$ states
associated to the $N^{\rm th}$ site (resp. rung) added to the 
chain ( resp. ladder). Eq. (\ref{1}) 
has to be supplemented with the initial data 
$| \beta \rangle_1$. 
The quantities $A_{\alpha, \beta}[s]$ are the variational
parameters of the MPM, and their determination is the central
problem one has to solve. This is done   by the standard
variational method. The important point
about eq.(\ref{1}) is that 
$A_{\alpha, \beta}[s]$ does not depend on $N$. Eq.(\ref{1})
is motivated by the truncation method used in the DMRG
where $A_{\alpha, \beta}[s]$  dependend on $N$,
i.e. $A_{\alpha, \beta}^{(N)} [s]$.
When $N$ is large enough one reaches a fixed point, i.e. 
$A_{\alpha, \beta}^{(N)}[s] \rightarrow A_{\alpha, \beta}[s]  $. 
In this manner the thermodynamic limit of  the
DMRG leads to a translational invariant MPM state.

The condition that both $|\alpha\rangle_N$ and
$|\beta\rangle_{N-1}$ form orthonormal basis imposes
a normalization condition on  $A_{\alpha, \beta}[s]$,

\begin{equation}
\sum_{\beta, s} A^*_{\alpha, \beta} [s] \; 
A_{\alpha',\beta}[s] 
= \delta_{\alpha ,\alpha'}
\label{2}
\end{equation}

It is interesting to count  how many variational 
parameters there are in (\ref{1}).
The quantities $A_{\alpha,\beta}[s]$ represent 
a total of $m^2 m^*$
variables. We shall assume that all of them 
may be non vanishing.
The normalization constraints (\ref{2}) represent a total of
$m + m(m-1)/2$ constraints ( $m$ coming from the diagonal terms
$\alpha = \alpha'$ and $m(m-1)/2$ 
coming from the off-diagonal ones). 
On the other hand one can rotate the basis of states 
$\{ |\alpha\rangle\}_{\alpha =1}^m$ 
by an element of the orthogonal group $O(m)$ reducing by 
$m(m-1)/2$ the number of independent MPM variables.
The total number of variational degrees of freedom, $N_A$,
is then  given by,

\begin{equation}
N_A = m^2 m^* - m - 2 m(m-1)/2 = m^2 (m^* -1) 
\label{3}
\end{equation}

We  show in appendix A that the set of 
 $A_{\alpha, \beta}[s]$
belongs, to  the
grassmanian manifold,

\begin{equation}
A \in \frac{O(m m^*)}{ O(m) \otimes O(m(m^* -1))}
\label{4}
\end{equation}

As an exercise one can check that the dimension of (\ref{4})
coincides with $N_A$ given in (\ref{3}).

In ref. \cite{OR} eq.(\ref{1}) is used to generate an
ansatz for the GS of
periodic chains. In this paper we
shall rather use this eq. to generate states with open
boundary conditions, in the spirit of the DMRG.
We shall show below that  the set $|\alpha\rangle$ 
correspond to 
ground states  with different boundary conditions.  
The use of open boundary conditions leads to a
simplification of the  MPM   
which is very close to the more abstract 
formalism proposed in \cite{Fan}.

\subsection*{ Correlators of local operators}

Let us use eq.(\ref{1})  
to compute the expectation values of local operators
in a recursive way. We shall first consider a local
operator ${\cal O}_n$ acting at the
position $n= 1, \dots, N$  of the lattice. It is easy
to get from (\ref{1}) the expectation value,

\[
_N\langle \alpha | {\cal O}_n | \alpha'\rangle_N
\]

\begin{equation}
= \left\{ \begin{array}{ll} 
\sum_{\beta \beta'} \;
{T}_{\alpha \alpha', \beta \beta'}
\; _{N-1}\langle \beta | {\cal O}_n | \beta'\rangle_{N-1}
& {\rm for} \;\; n < N \\
\sum_{\beta } 
\widehat{\cal O}_{\alpha \alpha', \beta \beta} & {\rm for}\;\; n=N
\end{array} \right.
\label{5}
\end{equation}

\noindent
where

\begin{equation}
T_{\alpha \alpha', \beta \beta'} = \sum_{s}
A^*_{\alpha, \beta} [s] \; A_{\alpha',\beta'}[s] 
\label{6} 
\end{equation}

\begin{equation}
\widehat{{\cal O}}_{\alpha \alpha', \beta \beta'} = \sum_{s s'}
A^*_{\alpha, \beta} [s] \;
\; A_{\alpha',\beta'}[s']  \;\;  \langle s  | {\cal O} | s' \rangle
\label{7}
\end{equation}

$T$ can be identified with $\widehat{1}$. 
Eqs.(\ref{5},\ref{6},\ref{7}) suggest to interpret
the expectation value 
$_N\langle \alpha | {\cal O}_n | \alpha'\rangle_N$ as
a vector labeled by the pair $\alpha \alpha'$, in which case
$T$ and $\widehat{\cal O}$ become 
$m^2 \times m^2$ 
matrices. 
Upon iteration of (\ref{5}) one finds,

\begin{equation}
_N\langle \alpha | {\cal O}_n | \alpha'\rangle_N
= \sum_{\beta }\; \left( T^{N-n}  \; \widehat{\cal O} 
\right)_{\alpha \alpha', \beta \beta} 
\label{8}
\end{equation}

More generally, the expectation value of a product
of local operators is given by,

\begin{equation}
_N\langle \alpha | {\cal O}_{n_1}^{(1)} \; {\cal O}_{n_2}^{(2)}
\cdots {\cal O}_{n_r}^{(r)}
| \alpha'\rangle_N 
\label{9} 
\end{equation}

\[
= \sum_{\beta }\; \left( T^{N-n_1}  \; \widehat{\cal O}^{(1)}
T^{n_1 - n_2 -1} \widehat{\cal O}^{(2)} \cdots T^{n_{r-1}-n_{r}-1}
\widehat{\cal O}^{(r)} \right)_{\alpha \alpha', \beta \beta}  
\]

\noindent where 
$N \geq  n_1 > n_2 > \cdots > n_r \geq 1$.
The matrix $T$ plays a very important role
in the MPM.
Eqs (\ref{5},\ref{9}) imply that  $T$ 
behaves  as a shift operator by one lattice space.
The basic properties of $T$ follow from 
the normalization condition (\ref{2}) which can be expressed
as,

\begin{equation}
\sum_{\beta} T_{\alpha \alpha', \beta \beta} 
= \delta_{\alpha, \alpha'}
\label{10}
\end{equation}

\noindent which implies that
$T$ has a eigenvalue equal to 1. 
Let us call  $|v\rangle$ 
the right eigenvector
corresponding to this eigenvalue. Eq.(\ref{10})
can be written 
in matrix notation as,

\begin{equation}
T \; |v \rangle = |v \rangle , \;\,\,
v_{\alpha \alpha'} = \delta_{\alpha \alpha'}
\label{11}
\end{equation}

On the other hand,  let $\langle \rho|$  denote the 
left  eigenvector of $T$ corresponding to the eigenvalue
1, i.e.

\begin{equation}
\langle \rho | T = \langle \rho | \leftrightarrow 
\sum_{\alpha \alpha'} \rho_{\alpha \alpha'} 
T_{\alpha \alpha', \beta \beta'} = \rho_{\beta \beta'}
\label{12}
\end{equation}

A convenient normalization of $\langle \rho|$ is
given by

\begin{equation}
 \langle \rho | v \rangle = 1\leftrightarrow
\sum_{\alpha} \; \rho_{\alpha \alpha} = 1
\label{13}
\end{equation}

For later use we shall diagonalize $T$ as follows,

\begin{equation}
T = \sum_p \; x_p |v_p\rangle \langle \rho_p|, \; \,\,
\langle \rho_p| v_{p'} \rangle = \delta_{p p'}
\label{14}
\end{equation}

\noindent where $|v_p\rangle$ and $\langle \rho_p|$ are
the right and left eigenvectors of $T$ with
eigenvalue $x_p \; \;  ( x_1 =1, \;  |v_1 \rangle =|v\rangle, 
\; \langle \rho_1| = \langle \rho |)$.
As a matter of fact all the remaining 
eigenvalues of $T$ are
less than one, i.e. $|x_p| < 1 \;\; \forall p\neq 1$.

In the limit $N\rightarrow \infty$  one gets,

\begin{eqnarray}
& \lim_{N \rightarrow \infty} \langle \alpha | 
{\cal O}_{n_1}^{(1)} \; {\cal O}_{n_2}^{(2)}
\cdots {\cal O}_{n_r}^{(r)}
| \alpha' \rangle_N & \label{15} \\
& =\; \delta_{\alpha \alpha'} \; \langle \rho |
\; \widehat{\cal O}^{(1)}
T^{n_1 - n_2 -1} \widehat{\cal O}^{(2)} \cdots T^{n_{r-1}-n_{r}-1}
\widehat{\cal O}^{(r)} | v \rangle & \nonumber  
\end{eqnarray}

The delta function on the r.h.s. of this eq. 
means that the local  operators $\widehat{\cal O}^{(n)}$
acting in the bulk, do not modify the 
boundary conditions associated to the 
various choices of $\alpha$. 
 
Assuming that $T$ is invertible, one can rewrite
eq.(\ref{15}) in the following manner,

\begin{eqnarray}
&  {\lim_{ N \rightarrow \infty}} _N \langle
\alpha | {\cal O}_{n_1}^{(1)} \; {\cal O}_{n_2}^{(2)}
\cdots {\cal O}_{n_r}^{(r)}
| \alpha' \rangle_N & \label{16} \\
& = \; \delta_{\alpha \alpha'} \; \langle \rho |
\; \tilde{\cal O}^{(1)}(n_1)
\tilde{\cal O}^{(2)}(n_2) \cdots 
\tilde{\cal O}^{(r)}(n_r) | v \rangle & \nonumber  
\end{eqnarray}

\noindent
where $\tilde{\cal O}(n)$ is defined as

\begin{equation}
\tilde{\cal O}(n) = T^{-n-1} \; \widehat{\cal O}\; T^n 
\label{17}
\end{equation}

Observe that $\tilde{ 1} = 1$. 
The r.h.s. of (\ref{16}) is nothing but  a spatial
ordered product of local operators $\tilde{\cal O}(n)$,
which is reminiscent of the radial ordered product
that appears in Conformal Field Theory. This connection
supports the interpretation of $T$ as an euclidean
version of the shift operator. Under this viewpoint 
the states $|v\rangle$ and $\langle \rho|$ appear
as incoming $|0\rangle$ and outgoing vacua $\langle 0|$
that are left invariant by the shift operator $T$.

We have shown above that the MPM 
leads in the thermodynamic limit  to a 
sort of discretized field theory
characterized by a shift or spatial transfer
operator $T$ and local operators $\tilde{\cal O}(n)$.
We can now try to exploit these interpretation to
extract some physical quantities.

First of all let us consider the correlator
of two operators ${\cal O}^{(1)}(n_1)$ and ${\cal O}^{(2)}(n_2)$. 
From
(\ref{14},\ref{15}) one has,

\begin{eqnarray}
&  \langle \rho |
\; \widehat{\cal O}^{(1)} \; T^{n_1-n_2 -1}
\widehat{\cal O}^{(2)}  | v \rangle  & \label{18} \\
& = \sum_p x_p^{n_1 -n_2 -1}
\; \langle \rho| \widehat{\cal O}^{(1)} |v_p \rangle \;
\langle \rho_p| \widehat{\cal O}^{(2)} |v \rangle & \nonumber  
\end{eqnarray}

In the limit when $|n_1 - n_2| >>1$ the sum over
$p$ is dominated by the highest  eigenvalue $|x_p|$ 
of $T$ for which
the corresponding matrix elements  
$\langle \rho| \widehat{\cal O}^{(1)} |v_p \rangle $ 
and 
$\langle \rho_p| \widehat{\cal O}^{(2)} |v \rangle $ are 
non zero. If $x_p < 1$ one gets a finite
correlation lenght  $\xi$ given by the formula,

\begin{equation}
\xi= -1/{\rm ln} |x_p|
\label{19}
\end{equation}

In the case where  $\widehat{\cal O}^{(1)}$ and  
$\widehat{\cal O}^{(2)}$ are both 
the spin operator ${\bf S}$, it turns out
that the matrix element $\langle \rho | \widehat{ {\bf S}} 
| v \rangle$
vanishes, and hence the spin-spin correlator  is short
ranged with a finite spin correlation length $\xi$
given by the
formula (\ref{19}) with $|x_p| < 1$.
The  finiteness of $\xi$ does indeed occur for MP
ansatzs which preserve the rotational invariance. 
However if the latter is broken, as in a Neel like
state, then $\xi$ may become infinite.

In section III we shall give a formula  
to compute $\xi$  in the case
of rotational invariant MP ansatzs.

Another
interesting application of (\ref{16}) 
is provided by
the computation of the 
string order parameter.

\subsection*{String order parameter}

A spin 1 chain has a long range  topological order (LRTO)
characterized by a non vanishing
value
of a non local operator $g(\infty)$  defined
as follows \cite{dNR},

\begin{eqnarray}
& g(\infty) = \lim_{\ell \rightarrow \infty} \; g(\ell),& \label{20} \\
&g(\ell) = \langle S^z(\ell) \; \prod_{k=1}^{\ell-1} 
e^{ \pi {\rm i} S^z(k)} \;  S^z(0) \; \rangle &
\nonumber
\end{eqnarray}

The AKLT state has $g_{AKLT}(\infty) =- (2/3)^2$, while the
spin 1 antiferromagnetic spin chain has 
$g(\infty) = -0.374325$ \cite{WH}. 
 From eq.(\ref{15}) we deduce the following expression
for (\ref{20}),

\begin{equation}
g(\ell) = \langle \rho| \widehat{S^z} \left( \widehat{e^{ {\rm i} 
\pi S^z } } \right)^{\ell -1} \; \widehat{S^z}| v \rangle 
\label{21}
\end{equation}

In appendix C we show that the operator 
$\widehat{e^{ {\rm i} \pi S^z } }$ has an eigenvalue 
equal to 1. Denoting by $|v^{\rm st} \rangle$
and $\langle \rho^{\rm st}|$ the associated 
right and left eigenvectors 
we obtain the following expression 
for the string order parameter

\begin{equation}
g(\infty) = \langle \rho|\; \widehat{S^z} \; |v^{\rm st}
\rangle \;  \langle \rho^{\rm st}|\; \widehat{S^z} \; 
|v \rangle 
\label{22}
\end{equation}

\noindent which  suggests that $g(\infty)$  measures 
a sort  of off-diagonal order.

For antiferromagnetic spin 1 ladders we shall see that the LRTO
disappears  and that the correlator (\ref{20})
is short ranged with a finite 
correlation length $\xi^{\rm st}$.

\subsection*{Ground state energy density}

Let us suppose we have a translational invariant Hamiltonian
of the form,

\begin{equation}
H_N = \sum_{n=1}^N
h^{(1)}_n + \sum_{n=1}^{N-1} \; h^{(2)}_{n,n+1}
\label{23}
\end{equation}

\noindent where $h^{(1)}$   is an on site (rung) operator 
while $h^{(2)}$ couples two nearest neighbour sites (rungs).
We define the expectation value,

\begin{equation}
E_{\alpha \alpha'}^{N} = _N\langle \alpha| H_N |\alpha' \rangle_N
\label{24}
\end{equation}

\noindent 
which can be computed 
recursively. From eqs.(\ref{1},\ref{5})
one gets

\begin{equation}
E^{N}_{\alpha  \alpha'} = 
\sum_{\beta \beta'} \; 
T_{\alpha \alpha', \beta \beta'} \; E^{N-1}_{\beta  \beta'}
+ \sum_{\beta} \widehat{h}_{\alpha \alpha',\beta \beta} ,
\,\,( N \geq 2 )
\label{25}
\end{equation}

\noindent where $\widehat{h} =  
\widehat{h}^{(1)} + \widehat{h}^{(2)}$. The hated
representation of the site hamiltonian $h^{(1)}$ is given by 
eq(\ref{7}), while the hated representation of the 
hamiltonian $h^{(2)}$ is given by,

\begin{eqnarray}
& \widehat{h}_{\alpha \alpha', \beta \beta'}^{(2)} =  
\sum_{\gamma \gamma' s's}  \;
_{N-1,N}\langle s_{2} s_{1}| h^{(2)}_{N-1,N} | 
s'_{1} s'_{2} \rangle_{N,N-1} & \label{26} \\
& \times \;
A^*_{\alpha, \gamma}[s_{1}] A^*_{\gamma, \beta}[s_{2}] 
A_{\alpha', \gamma'}[s'_{1}] A_{\gamma', \beta'}[s'_{2}] 
& \nonumber  
\end{eqnarray}

It should be clear from eqs.(\ref{7}, \ref{26}) which is the
hated  representative 
of an operator involving an arbitrary
number of sites.
Eq.(\ref{25}) can be conveniently written in matrix notation
as

\begin{equation}
|E^{N} \rangle = T \; |E^{N-1} \rangle + \widehat{h} \; |v \rangle,
\;\;  (N \geq 2)
\label{27}
\end{equation}

\noindent where $|E^{N} \rangle $ is a vector with components
$E^{N}_{\alpha  \alpha'}$.
Iterating (\ref{27})  one gets

\begin{equation}
|E^{N} \rangle = ( 1 + T + T^2 + \cdots + T^{N-2} )\;  \widehat{h}\; 
|v \rangle
+ T^{N-1} \; |E^{1} \rangle 
\label{28}
\end{equation}

The geometric series in $T$ can be sumed up and due
to the eigenvalue equal to 1 
it contributes
a term proportional to $N$, i.e.

\begin{equation}
\lim_{N \rightarrow \infty} \frac{1}{N} |E^{N}\rangle
= e_{\infty} \; |v \rangle
\label{29}
\end{equation}

\noindent
This eq. implies that all the states 
$|\alpha \rangle_N$ have the same energy density
in the thermodynamic limit, i.e.  
$ E_{\alpha \alpha'}^{N} = \delta_{\alpha \alpha'} \;
e_{\infty}$. 
Hence  $e_{\infty}$ can be identified with the
GS energy per site for chains or per rung for ladders
and it is 
given by,

\begin{equation}
e_\infty =  \langle \rho|\; \widehat{h}\;  
| v \rangle = \sum_{\alpha
\alpha' \beta } \rho_{\alpha \alpha'} \; 
\widehat{h}_{\alpha \alpha',
\beta \beta}
\label{30}
\end{equation}

This is the quantity one has to minimize 
respect to the MPM parameters.

The formalism presented above is closely
related to the DMRG. Eventhough this relation
is not the main subject of this paper we shall
make some remarks ( see \cite{D}).

\subsection*{MPM versus the DMRG}

Let us suppose that we diagonalize  
$\rho$, as a $m \times m$ matrix, denoting  its eigenvalues
as  $w^2_{\alpha}$, i.e.

\begin{equation}
\rho_{\alpha \alpha'} = w^2_{\alpha} \, 
\delta_{\alpha \alpha'} 
\label{34}
\end{equation}

The eigenvalue eq.(\ref{12}) becomes then,

\begin{equation}
\sum_{\alpha s} \;  w^2_{\alpha}  A_{\alpha \beta}[s]
\; A^*_{\alpha \beta'}[s]  = \;  \delta_{\beta \beta'} 
w^2_{\beta}
\label{35}
\end{equation}

There is a close analogy between 
 eqs(\ref{2}) and (\ref{35}),
except for the fact that the order of the labels
is exchanged. 
Given  the tensor product decomposition
$s \otimes \beta
\rightarrow \alpha$, we shall assume that 
one can reverse the order
between the states 
$\alpha $ and $\beta$ in terms of 
``charge conjugate states'' $\alpha^c$ and $\beta^c$, 
as follows : $s \otimes \alpha^c \rightarrow \beta^c$.
For example the charge conjugate of a state with spin 
$M$ is another state with spin $-M$.
Using this concept we can impose the following 
symmetry condition \cite{NO},

\begin{equation}
w_{\alpha}  A_{\alpha \beta}[s] = \pm 
w_{\beta^c}  A_{\beta^c \alpha^c}[s], \;\; 
w_{\beta^c}= w_{\beta} 
\label{36}
\end{equation}

\noindent which leads to the equivalence between 
eqs. (\ref{2}) and (\ref{35}). 

The relation between the MPM and the DMRG 
is made clear by the construction of  the GS of the superblock
$B_N  \bullet B^R_{N} $ in the following way,

\begin{eqnarray}
& | \psi_0 \rangle = \sum
\psi_{\alpha s \beta } \;\;  |\alpha^R \; \rangle \otimes
| s \rangle \otimes |\beta \rangle & \nonumber \\
& \psi_{\alpha s \beta } = w_\alpha \; A_{\alpha \beta}[s]
& \label{37} 
\end{eqnarray}

The density matrix that induces $\psi_{\alpha s \beta }$ 
on the block $B$, and which  is obtained by tracing
over the states in $ \bullet B^R_{N} $, coincides 
with $\rho_{\alpha \alpha'}= w^2_{\alpha} \, \delta_{\alpha \alpha'}$. 

Condition (\ref{36}) guarantees  that 
$|\psi_0\rangle$ is a state invariant under the parity
transformation that interchanges the blocks $B$  and $B^R_{N} $,
while leaving invariant the site $\bullet$. 

It is interesting to observe that the MPM leads to a superblock
of the form $B_N  \bullet B^R_{N} $, rather than   to the standard
superblock $B_N  \bullet \bullet B^R_{N} $ \cite{D}.

\section*{III) The MPM applied to spin ladders}

In this section we shall apply the MPM
to the 2-leg spin ladder
with a spin $S$ at each site of the chain. 
The collection $|\alpha \rangle_N$ will be given 
by  the set $|J M \rangle_N \;  (J \leq  J_{\rm max}$)
of states with total spin $J$ and third component $M$.
For the sake of simplicity we have only considered
one state per angular momenta $J$ and $M$. 
This will allow us to show more 
clearly the analytic structure of the MPM, which 
can later on be numerically improved by considering multiplicity.
This has  already been 
done in the case of spin chains in references \cite{OR,D}.

The states added at each step of the MPM
are the ones that appear in the tensor product
decomposition of two spin $S$ irreps, i.e.
$S \otimes S = 0 \oplus 1 \oplus \cdots \oplus 2 S$. 
These states
are labelled by $|\lambda \mu\rangle$ where $\lambda = 0, \cdots, 2 S$
is the total spin and 
$\mu = - \lambda, \dots , \lambda$ is its third component.

Using these notations we propose the following 
recurrence relation for the states 
$|J M\rangle_N$,

\begin{equation}
|J_1 M_1\rangle_N = \sum_{\lambda J_2} \;
A_{J_1 J_2}^\lambda \; |(\lambda J_2), J_1 M_1 \rangle_N 
\label{a1} 
\end{equation}

\noindent where  

\begin{equation}
|(\lambda J_2), J_1 M_1 \rangle_N 
\label{a2} 
\end{equation}

\[
= \sum_{\mu} \; \langle \lambda \mu, J_2 M_2|
\lambda J_2 ,J_1 M_1 \rangle \;
\; |\lambda \mu\rangle_N \otimes |J_2 M_2 \rangle_{N-1} 
\]

\noindent In (\ref{a2}) the quantity 
$\langle \lambda \mu, J_2 M_2|\lambda J_2, J_1 M_1 \rangle$ 
is the
Clebsch-Gordan coefficient corresponding to the
decomposition $\lambda \otimes J_2 \rightarrow J_1$.
Comparing eqs.(\ref{1}) and (\ref{a1}) we obtain
the following relation between the symbols 
$A_{J_1 M_1, J_2 M_2}[\lambda \mu]$ and the rotational 
invariant symbols $A_{J_1 J_2}^\lambda$,

\begin{equation}
A_{J_1 M_1, J_2 M_2}[\lambda \mu]
= A_{J_1 J_2}^\lambda \;
\langle \lambda \mu, J_2 M_2|\lambda J_2, J_1 M_1 \rangle
\label{a3}
\end{equation}

The use of rotational
invariant basis reduces considerably the number of
independent variational parameters and consequently
increases the power of the MPM \cite{OR,D}.

The variational parameters $A_{J_1 J_2}^\lambda$ are subject 
to the CG condition,

\begin{equation}
A_{J_1 J_2}^\lambda = 0 \;\;{\rm unless }\;\; |\lambda - J_2|
\leq J_1 \leq |\lambda + J_2|
\label{a4}
\end{equation}

Using  (\ref{a3}) and  
the orthogonality properties of the CG coefficients,  
the normalization conditions (\ref{2}) become,

\begin{equation}
\sum_{\lambda, J_2} \; |A^\lambda_{J_1 J_2}|^2 = 1 ,  \; \; \; \forall \; J_1
\label{a5}
\end{equation}

At this point we can just take eq.(\ref{a3}) and plug it into 
the corresponding formulas of section II in order to derive 
expectation values, the GS energy density, etc in terms 
of $A_{J_1 J_2}^\lambda$. There is however 
a more  efficient way to do this by using group theory.
The application of the Wigner-Eckart theorem will allow
us to  express all the results in terms of reduced matrix elements
of the operators involved as well as the $6j$-symbols. 
In our derivations we shall  follow the same steps as
in section II, leaving the technical details to appendix B.

\subsection*{Correlators of Invariant Tensors}

Let us denote by ${\cal O}^{(k)}$ an irreducible tensor of 
total angular momentum $k$, whose components 
are labeled by  ${\cal O}^{(k)}_M , M=- k, \dots, k$. 
The spin operators
${\bf S}$ correspond to
 $k=1$. Let us suppose we have 
two irreducible tensors  with the same total angular momenta $k$,
${\cal O}^{(k,A)}(n)$
and  ${\cal O}^{(k,B)}(m)$, acting at the positions 
$n$ and $m$ ( $N \geq n >m \geq 1 $) of the ladder. 
The scalar product of these two operators is defined
as

\begin{equation}
{\cal O}^{(k,A)}(n) \cdot {\cal O}^{(k,B)}(m)
\label{a6}
\end{equation}

\[
= \sum_{M = -k}^{k} \; {( -1)}^{-M} 
{\cal O}^{(k,A)}_{M}(n) \cdot {\cal O}^{(k,B)}_{-M}(m)
\]

The basic result we derive in appendix B 
is,

\begin{equation}
_N\langle J_1 M| \;
{\cal O}^{(k,A)}(n) \cdot {\cal O}^{(k,B)}(m)
\; | J_1 M \rangle_N 
\label{a7} 
\end{equation}

\[=
\sum_{J_2, \dots, J_7} \; 
T^{N-n}_{J_1 J_2} \; 
\widehat{\cal O}^{(k,A)}_{J_2, J_3 J_4}\; 
\left(T_{k}^{n-m-1}\right)_{J_3 J_4, J_5 J_6} \;
\widehat{{\cal O}}^{(k,B)}_{J_5 J_6, J_7}
\]

\noindent where  
$ T $ and  
$ T^{(k)}$ are defined as,

\begin{eqnarray}
& T_{J_1, J_2} = \sum_\lambda \; 
\left( A_{J_1 J_2}^\lambda \right)^*
A_{J_1 J_2}^\lambda &
\label{a8}
\end{eqnarray}

\begin{eqnarray}
& \left(T_k \right)_{J_1 J_2, J_3 J_4 } 
= \sum_\lambda \; 
\left( A_{J_1 J_3}^\lambda \right)^*
A_{J_2 J_4}^\lambda & \label{a9} \\
& \times \; 
(-1)^{\lambda + k + J_1 + J_4} \sqrt{(2 J_1 +1) (2 J_2 +1)}
\; \left\{ \begin{array}{ccc} 
J_3 &  J_1  & \lambda \\
J_2 &  J_4  & k \end{array}
\right\} &
\nonumber 
\end{eqnarray}

\noindent while $\widehat{\cal O}^{(k,B)}$ and $\widehat{\cal O}^{(k,B)}$
are defined in  Appendix B.

Eq.(\ref{a7}) is the invariant version of (\ref{9})
involving  only two operators. In order to obtain the
thermodynamic properties of (\ref{a7}) we use 
the properties of the transfer operator $T$.
The normalization conditions (\ref{a5}) imply the
following conditions on $T$,

\begin{equation}
\sum_{J_2} \; T_{J_1 , J_2} = 1 , \;\; \forall \; J_1
\label{a12}
\end{equation}

Let us call  $\rho_J$ the
left eigenvector of $T_{J_1, J_2}$ with eigenvalue 1, i.e.

\begin{equation}
\sum_{J_1} \; \rho_{J_1} \; 
T_{J_1 , J_2} = \rho_{J_2} 
\label{a13}
\end{equation}

Using eqs.(\ref{a12}, \ref{a13}) into (\ref{a7}) and taking 
$N >>1$ we get

\begin{eqnarray}
& \lim_{N\rightarrow \infty} \; 
 _N\langle J_1 M| \;
{\cal O}^{(k,A)}(n) \cdot {\cal O}^{(k,B)}(m)
\; | J_1 M \rangle_N & 
\label{a14} \\        
& = \langle \rho |  \; 
 \widehat{\cal O}^{(k,A)}\; T_{k}^{n-m-1} \;
\widehat{{\cal O}}^{(k,B)} |v \rangle  & \nonumber  
\end{eqnarray}

\noindent where we use a matrix notation in 
$J$-space with the convention $v_J = 1, \;\forall J$.
 From eq.(\ref{a14}) we deduce that 
the correlation length associated to the 
scalar product of two irreducible operators
with angular momentum $k$, is given by the highest eigenvalue
of the matrix $T_k$ defined in (\ref{a9}). 
The spin-spin correlation length is obtained by looking
at the highest absolute eigenvalue of $T_1$.

\subsection*{Ground state energy density}

The Hamiltonian of the 2-leg 
ladder has the form proposed in (\ref{23})
where $h^{(1)}$ is the rung Hamiltonian and $h^{(2)}$ is the leg
Hamiltonian,

\begin{eqnarray}
& h^{(1)}_n = J_\perp \; {\bf S}_1(n) \cdot {\bf S}_2(n) &
\label{a15} \\
&  h^{(2)}_{n, n+1} = J_\parallel \;
\left( {\bf S}_1(n) \cdot {\bf S}_1(n+1) +
{\bf S}_2(n) \cdot {\bf S}_2(n+ 1) \right) &
\label{a16} 
\end{eqnarray}

\noindent 
${\bf S}_a(n)$ is a spin $S$ operator
acting on  the $n = 1, \cdots, N $ rung and the $a = 1,2$ leg 
of the ladder.

As in (\ref{24}) we define the expectation value of the
ladder Hamiltonian,

\begin{equation}
E^{N}_J = _N\langle J M | H_N | J M \rangle_N 
\label{a17}
\end{equation}

Using  (\ref{a1}) we find

\begin{equation}
E^{N}_{J_1} = \sum_{J_2} \; \left( 
 T_{J_1, J_2} \; E^{N-1}_{J_2} + \widehat{h}_{J_1, J_2}
\right) , \;\;(N \geq 2)
\label{a18}
\end{equation}

\noindent where
$ \widehat{h} = \widehat{h}^{(1)} +  \widehat{h}^{(2)} $
($\widehat{h}^{(1)}$ and   $\widehat{h}^{(2)} $ can be found
in appendix B).

Iterating eq.(\ref{a18}) and using the properties 
of the matrix $T$ we can inmediately get the
large $N$ limit of the energy (\ref{a17}),

\begin{equation}
\lim_{N \rightarrow \infty} \frac{1}{N} 
E^{N}_{J} = e_\infty , \;\; \forall \; J 
\label{a21}
\end{equation}

\noindent
where the GS energy density is given by,

\begin{equation}
e_{\infty} = \langle \rho| \; \widehat{h} \;| v \rangle
= \sum_{J_1, J_2} \rho_{J_1} \; \widehat{h}_{J_1 J_2}
\label{a22}
\end{equation}

At this point 
let us summarize the main steps
of the MP algorithm hereby proposed,

\begin{itemize}

\item Solve the normalization conditions (\ref{a5})
expressing $A^{\lambda}_{J_1 J_2}$ in terms of a set
of linearly independent variational parameters.

\item Find the eigenvector  $\rho_J$ of the matrix $T$.

\item Minimize the GS energy density (\ref{a22})
with respect to the independent variational parameters.

\end{itemize}

We will now comment on how these three steps can be implemented.

\subsection*{Solution of the normalization conditions}

We shall suppose in the rest of the paper that
the parameters  
$A^{\lambda}_{J_1 J_2}$ are all real. 
Hence the normalization conditions

\begin{equation}
\sum_{\lambda,  J_2} \; (A^\lambda_{J_1 J_2})^2 = 1 ,
\; \; \; \forall \; J_1
\label{a23}
\end{equation}

\noindent imply that the set
$\{ A^{\lambda}_{J_1 J_2} \} $ for $J_1$ fixed 
are the coordinates of a sphere whose dimension depends
on the allowed values of $J$ and the CG conditions (\ref{a4}).
Let us call
$A^{{\rm max} }_{J_1}$ the highest coordinate,
in absolute value, i.e.

\begin{equation}
A^{{\rm max} }_{J_1} = A^{\lambda_0}_{J_1 L_0} \; {\rm such} \;{\rm that}\;
|A^{\lambda_0}_{J_1 L_0}| \geq  
|A^{\lambda}_{J_1 J_2}| \;\;   \forall \; \lambda, J_2
\label{a24}
\end{equation}

If $A^{{\rm max} }_{J_1} >0 $( resp. $A^{{\rm max} }_{J_1} <0)$
we can think of it as the north ( resp. south) pole
of a sphere, whose neaghbourhood can be described
by the stereographic coordinates,

\begin{equation}
x^{\lambda}_{J_1 J_2} = 
A^{\lambda}_{J_1 J_2}/ A^{{\rm max} }_{J_1}, 
\; \;\, |x^{\lambda}_{J_1 J_2}| \leq 1
\label{a25}
\end{equation}

Notice that  
$x^{\lambda_0}_{J_1 L_0} =1$. The remaining coordinates  
are the independent
variational parameters used in the minimization
of the GS energy. The solution of the constraint (\ref{a23})
finally reads,

\begin{equation}
A^{\lambda}_{J_1 J_2} = \epsilon_{J_1} \; 
x^{\lambda}_{J_1 J_2} \left( \sum_{\lambda', J_2'}  
\; \; \left( x^{\lambda'}_{J_1 J'_2} \right)^2 \right)^{-1/2}, \; \;
\epsilon_{J} = \pm 1
\label{a26}
\end{equation}

\noindent where $\epsilon_{J_1} = 1 (-1)$ corresponding 
to the north ( south ) pole of the above mentioned sphere.

\subsection*{Determination of $ \rho_J$}

The solution of the eigenvalue problem of 
eq.(\ref{a13}) can be done 
numerically. However for a ladder with spin $S=1/2$ it can also be solved
analytically which will allow us to make some  considerations
on the nature of $\rho_J$.
In the case where $S=1/2$ the allowed values for $\lambda$ are 0 and 1.
Hence the unique non-vanishing entries of 
$A^{\lambda}_{J_1 J_2}$ are  $A^{0}_{J J}, A^{1}_{J J},
A^{1}_{J J+1}$ and  $A^{1}_{J J-1}$. Similarly
from eq.(\ref{a8}) the non-vanishing entries of $T$ are
$T_{J,J}, T_{J,J+1}$ and $ T_{J,J-1}$. The set of eqs. we have therefore
to solve read explicitely as,

\begin{eqnarray}
& T_{J,J}+ T_{J,J+1}+  T_{J,J-1} = 1 & \nonumber \\
& \rho_{J} \; T_{J,J}+ \rho_{J+1} \; T_{J+1,J}+ 
\rho_{J-1}  T_{J-1,J} = \rho_J  & \label{a27} \\
& \sum_{J=0}^{J_{\rm max}} \; \rho_j = 1 & \nonumber 
\end{eqnarray}

The solution of these eqs. is given by,

\begin{equation}
\rho_J = \frac{u_J}{ \sum_{L}  u_L }
\label{a28}
\end{equation}

\noindent where

\begin{equation}
u_0 = 1, \;\; 
u_J = \prod_{L=0}^{{J}}\; \frac{T_{L,L+1}}{ T_{L+1,L}}\,\,\; ( J>0)
\label{a29}
\end{equation}

\noindent we are assuming that $J = 0, \cdots , J_{\rm max}$.

Eqs.(\ref{a28}, \ref{a29}) imply 
that $\rho_J$ is always positive, in agreement  with the 
Perron-Frobenius theorem applied to the matrix $T$, whose
entries are all non-negative. 
In ref.\cite{D} it was shown that the values of
$\rho_J$ are 
intimately related to 
the eigenvalues of the density matrix
that appear in the DMRG. These and other facts suggest 
that the MPM is in fact equivalent to the DMRG, specially
when the number of states kept $m$ becomes large.

This completes  the presentation of the formalism.

\section*{IV) Numerical Results}

In this section we shall apply the MPM to 
five different spin ladders, corresponding to different
choices of the spin $S$ and signs of the coupling constants
$J_\parallel$ and $J_\perp$. We shall denote every of these
ladders as $AA_S, AF_S$, and $FA_S$ where $A$ and $F$ stands
for antiferromagnetic or 
ferromagnetic couplings. Thus for example $AF_S$ 
is a spin S ladder with  antiferromagnetic
couplings   
along the legs and  ferromagnetic couplings 
along the rungs. 
With these notations we will study below the following cases:
$AA_{1/2}$, $AF_{1/2}$, $FA_{1/2}$, $AA_{1}$ and 
$AA_{3/2}$.
Within each case we will highlight 
a particular aspect, which the MPM
helps to clarify.

\subsection*{$AA_{1/2}$-ladder: The dimer-RVB state}

This is the most studied spin ladder. Its properties are well
known and can be summarized as follows. In the weak coupling regime,
i.e. $J_\perp << J_\parallel$, 
the  gapless  spin 1/2 chains 
become massive by the interchain coupling which is a relevant
operator of dimension 
1 \cite{Schu,SM,SNT}. The magnitude of the gap is 
proportional to $J_\perp$. In the intermediate coupling regime, 
i.e. $J_\perp \simeq J_\parallel$ the spin ladder can be mapped into
the $O(3)$ non linear sigma 
model (NLSM) with no topological term \cite{Sen,Sie,M}.
This model is known to have a spin gap. From numerical studies
the magnitude of the spin gap $\Delta$ and the spin correlation
length, in the isotropic case
$J_\perp = J_\parallel=J $, are given by  
$\Delta = 0.502 J$ and $\xi = 3.2$ respectively \cite{DRS,WNS,gap1,gap2}. 
In the strong coupling regime 
$J_\perp >> J_\parallel$, the most appropiate physical  
picture of the GS and excitations is given by 
the RVB scenario proposed in \cite{WNS}, and supported by
DMRG \cite{WNS}, mean field \cite{mean}
and variational calculations \cite{SMA}.
In the latter work a recurrent variational ansatz (RVA)  was proposed
to generate the dimer-RVB and generalizations of it. The RVA
method is a MPM based 
on $2^{\rm nd}$  and higher
order recurrent relations, while 
the standard MPM is based on a $1^{\rm st}$ order relation.
We shall see below that the MPM applied to ladders 
essentially  contains the RVA, and that the numerical results are
improved.

Let us first consider the case where the MPM states
$|J M\rangle_N$ 
are choosen to be 
a singlet and a triplet, i.e.. $J = 0$ and 1.  
In this case eq. (\ref{a1}) is depicted in fig.~1. 
There are a priori 5 non vanishing MP parameters 
subjected  to 2 normalization constraints,
leaving  a total of 3 independent parameters.

Fig.2 shows $A_{J_1 J_2}^\lambda$ as functions of
$x= J_\parallel/ J_\perp$.
In the whole range of coupling constants
the most important amplitudes are $A^0_{0 0}$ and
$A^1_{1 0}$ while $A^1_{1 1}$ is essentially zero.
The latter amplitude correspond to having 
only a single bond among two rungs ( see fig. 1),
which is forbidden in  the dimer-RVB  
picture of refs.\cite{WNS} and \cite{SMA}.

In table 1  we show the GS energy density obtained 
with the MPM for $J_{\rm max} = 1$ and 2, together
with the RVA, mean field and Lanczos results. 
In table 2 we give the spin correlation length computed
with the MPM and the RVA.

There is an appreciable improvement
in the numerical
results  of the MPM respect
to the RVA,
specially for  the spin correlation length.

\begin{center}
\begin{tabular}{cccccc} 
\hline \hline
$J_\parallel/J_\perp$ & $J_{\rm max} =1$ &  $J_{\rm max} =2$   
& RVA & Mean Field & Lanczos\\  \hline \hline

$0.0$ & $0.375000$& $0.375000$ & $0.375000$ & $0.375000$  & $$ \\ \hline 

$0.2$& $0.383199$& $0.383199$& $0.383195$ & $0.382548$  & $$ \\ \hline 

$0.4$& $0.409607$& $0.409608$& $0.409442$ & $0.405430$  & $$ \\ \hline 

$0.6$& $0.453509$& $0.453513$& $0.45252$ & $0.442424$  & $$ \\ \hline 

$0.8$& $0.510504$& $0.510523$& $0.507909$ & $0.489552$  & $$ \\ \hline 

$1.0$&  $0.575924$& $0.575970$& $0.571314$ & $0.542848$  & $0.578$ \\ \hline 

$1.25$& $0.664776$& $0.664867$& $0.657551$&$0.614473$ &$0.6687$ \\ \hline 

$1.66$& $0.819656$& $0.819834$&  $0.808438$&$0.738360$ & $0.8333$ \\ \hline 

$2.5$& $1.152056$& $1.152416$& $1.13384$&$1.002856$ & $1.18$ \\ \hline 

$5.0$& $2.172002$& $2.172878$& $2.13608$ & $$  & $2.265$ \\ \hline \hline 
\end{tabular}
\end{center}
\begin{center}
Table 1. GS energy per site 
$- e_{\infty}/2 J_{\perp}$ of the ladder $AA_{1/2}$. 
The first two columns are
the MPM results. The RVA results are obtained with a third
order recursion formula \cite{SMA}. The mean field and Lanczos
results have been obtained in  references \cite{mean} and \cite{Lan}
respectively.   
\end{center}

\begin{center}
\begin{tabular}{cccc} 
\hline \hline
$J_\parallel/J_\perp$ &   $J_{\rm max} =1$ &  $J_{\rm max} =2$
& RVA  \\  \hline \hline

  $0.0$ & $0.00000$ & $0.0000$ & $0.000000$  \\ \hline 

  $0.2$  & $0.5300$& $0.5303$ & $0.437166$  \\ \hline 

  $0.4$ & $0.8057$& $0.8081$ & $0.608323$   \\ \hline 

  $0.6$ & $1.0652$ & $1.0740$ & $0.751286$   \\ \hline 

  $0.8$ & $1.2753$& $1.2945$ & $0.866958$  \\ \hline 

  $1.0$ & $1.4282$ & $1.4593$ & $0.959249$   \\ \hline 

  $1.25$ & $1.5572$ & $1.6018$ & $1.04877$   \\ \hline 

  $1.66$ & $1.6802$ & $1.7413$ & $1.15205$  \\ \hline 

  $2.5$ & $1.7903$ & $1.8698$ & $1.26951$  \\ \hline 

  $5.0$ & $1.8747$& $ 1.9711$  & $1.38532$   \\ \hline \hline 
\end{tabular}
\end{center}
\begin{center}
{Table 2} Spin correlation length of the ladder $AA_{1/2}$.
The first two columns are the MPM results. 
The RVA results are those of ref \cite{SMA}. 
\end{center}

\subsection*{$AF_{1/2}$-ladder:
Relation with  the spin 1 chain}

The ladder with magnetic structure $AF$
is interesting because it is intimately
related to the spin 1 chain \cite{Hi}. 
This relation can be clearly seen in 
the strong coupling limit $ -J_{\perp} >> J_\parallel$, since it
leads to an effective spin 1 on  every rung, 
which are effectively coupled antiferromagnetically along
the legs. The effective Hamiltonian can be derived from
(\ref{a15},\ref{a16}) and reads \cite{Hi},

\begin{equation}
H_{\rm eff}^{\rm ladder}= - \frac{1}{4} \;|J_{\perp}| \;N
+ \frac{1}{2} \; J_\parallel \; \sum_{n} \; {\bf S}_{\rm eff} \cdot
{\bf S}_{\rm eff} 
\label{b1}
\end{equation}

\noindent where ${\bf S}_{\rm eff}(n)= {\bf S}_1(n) + {\bf S}_2(n)$ 
is the spin 1 operator acting on  the $n^{\rm th}$ rung. 
The term proportional to $J_{\perp}$ comes from the rung
Hamiltonian when diagonalized in the spin 1 sector.
This eq. implies the following relation between the energies
per site of the $AF$-ladder and the spin 1 chain,

\begin{equation}
e_{\infty}^{AF} = - \frac{1}{8} \; | J_\perp | + \frac{1}{4}
\; J_\parallel \; e_{\infty}^{\rm eff}
 \label{b2}
\end{equation}

In table 3 we give the GS energies of the ladder parametrized
in terms
of the effective energy $e_{\infty}^{\rm eff}$. We also
give the spin correlation length.  
We have made  two choices  of MP states $|JM\rangle_N$.
One for which $J=0$ and 1 and the other for which  $J=1/2$ and 3/2.
The latter one corresponds to having a single spin 1/2 at
the boundary of the ladder. These two choices have
an analogue for the spin 1 chain. 
For  integer $J's$  the MP parameters do not vary in the
whole interval $0 < J_\parallel < 1.66$ and take the values,

\begin{eqnarray}
& A^0_{0 0 } = A^0_{1 1} = 0.000, A^1_{0 1} =  1.000 & \label{b3} \\
& A^1_{1 0}= -0.577, A^1_{1 1} = 0.816 & \nonumber 
\end{eqnarray}

For  half-integer $J's$  the MP parameters do not vary in the
whole interval $0 < J_\parallel < 5$ and take the values,

\begin{eqnarray}
& A^0_{\frac{1}{2} \frac{1}{2}} = 
A^0_{ \frac{3}{2} \frac{3}{2}} = 0.000, 
A^1_{ \frac{1}{2} \frac{1}{2}} = 0.989 & \label{b4} \\
& A^1_{\frac{1}{2} \frac{3}{2}} = 0.148, 
A^1_{ \frac{3}{2} \frac{1}{2}} = -0.953, 
A^1_{ \frac{3}{2} \frac{3}{2}} = -0.303 & \nonumber 
\end{eqnarray}

Note 
that for half integers $J$ there is one more variational parameter.
The most important amplitudes indeed correspond to the formation
of an AKLT state with  a single bond connecting every effective
spin 1.

\begin{center}
\begin{tabular}{ccccc}
\hline
$-J_\parallel/J_\perp$ & $-e_\infty^{\rm eff,int}$ & $
-e_\infty^{\rm eff, half}$ & $\xi^{\rm int}$ & $\xi^{\rm half}$ \\ 
\hline
\hline
(0.0, 1.66) & $1.333333$ & $1.399659$ & 0.910 & 2.5997  \\ 
\hline
$2.5$ & 1.363970  & $1.399659$ & 1.9682 & 2.5997 \\ 
\hline
$5.0$ &$1.498539$ & $1.399659$ & 1.9607 & 2.5997 \\
\hline
\end{tabular}
\end{center}
\begin{center}
Table 3.  The exact values of the GS energy density of a spin
1 chain and its correlation length are given by
$e_{\infty} = -1.4014845$ and $\xi= 6.03$ \cite{WH}.
\end{center}

The values
of $e_\infty^{\rm eff}$ and $\xi$ in the integer $J$ case 
coincide with those of the AKLT state, while the 
ones of the half-integer $J$ case coincide with 
those obtained
with a MPM method applied to the spin 1 chain \cite{D} 
where one keeps two MPM states with $J=1/2$ and 3/2.
These results provide additional support for  the equivalence
between the $AF_{1/2}$ ladder and the spin 1 chain in the strong 
and intermediate coupling regimes observed previously by
other authors \cite{Hi}.

\subsection*{$FA_{1/2}$-ladder}

In the strong coupling regime the ladders $FA_{1/2}$ and
$AF_{1/2}$ have similar GS energies and correlation lengths
( see tables 1, 2 and 4). The MP parameters display also a similar
behaviour although some of them  are interchanged 
( see figures 2 and 3). The physical reason of this is 
the common GS in the case where $J_\parallel = 0$,
given by the coherent superposition  of valence bonds 
in the rungs.

\begin{center}
\begin{tabular}{ccc}
\hline
$-J_\parallel/J_\perp$ & $-e_\infty/2 J_\perp$ & $\xi$ \\ 
\hline
\hline
0.0 & $0.375000$ & 0.0000  \\ 
\hline
$0.2$ & 0.381754  & 0.5140 \\ 
\hline
$0.4$ & 0.399295 & 0.7577 \\
\hline
$0.6$ & 0.424396 & 1.010 \\
\hline
$0.8 $ & 0.454891 & 1.277 \\
\hline
$1.0$ & 0.489324 & 1.554 \\
\hline
$1.25$ & 0.536374 & 1.895  \\
\hline
$1.66$ & 0.619895 & 2.381 \\
\hline
$2.5$ & 0.803434 & 2.992 \\
\hline 
$5.0$ & 1.376973 & 3.520 \\
\hline 
\end{tabular}
\end{center}
\begin{center}
Table 4. GS energy per site  and 
correlation length of the ladder $FA_{1/2}$.
\end{center}

The relation between $FA_{1/2}$ and
$AF_{1/2}$ is part of a more general relation 
involving  also the ladder $AA_{1/2}$ and can be established
by means  of a type of transformations called dualities in ref. 
\cite{dual}.

\subsection*{Duality properties of spin ladders}

On a  2-leg ladder 
one can define  3 types of dualities called $U, T $ and $S$, which 
mix or leave invariant 
the ladder's magnetic structures $AA, AF$ and $FA$ \cite{dual}.

The $U$ duality maps a Hamiltonian with couplings constants
$J_\parallel, J_\perp$ into a ladder with   couplings constants
$J^U_\parallel, J^U_\perp$ where,

\begin{eqnarray}
& J^U_\parallel = J_\parallel \; 
\langle {\bf S}_1(n) \cdot {\bf S}_1(n+1) \rangle/ 
\langle {\bf S}_1(n) \cdot {\bf S}_2(n+1) \rangle & \label{b5}   \\
& J^U_\perp  = J_\perp & \nonumber
\end{eqnarray}

Under  $U$ the leg-bonds are transformed into diagonal ones
while the rung-bonds are left invariant. The signs
of $\langle {\bf S}_1(n) \cdot {\bf S}_1(n+1) \rangle $ and 
$\langle {\bf S}_1(n) \cdot {\bf S}_2(n+1) \rangle $ are determined
by those of $J_\parallel$ and $J_\perp$ respectively. 
Thus  $U$ acts on the magnetic structures as follows,

\begin{equation}
\begin{array}{ccc} 
AA & \stackrel{U}{\rightarrow} & FA \\
J_\parallel( AA) > 0 & {\rightarrow} & J_\parallel^U(FA) <0 \\
J_\perp(AA) > 0 & {\rightarrow} &  J_\perp^U(FA) >0  
\end{array}
\label{b6}
\end{equation}

\begin{equation}
\begin{array}{ccc} 
AF & \stackrel{U}{\rightarrow} & AF \\
J_\parallel( AF) > 0 & {\rightarrow} & J_\parallel^U(AF) > 0 \\
J_\perp(AF) < 0 & {\rightarrow} &  J_\perp^U(AF) < 0  
\end{array}
\label{b7}
\end{equation}

In fig.4 we show $J_\parallel^U(FA)$ and  $J_\parallel^U(AF)$ 
as functions of $J_\parallel(AA)$ and $J_\parallel(AF)$ respectively.

The GS energy density of the ladder with coupling constants 
$J^U_\parallel, J^U_\perp$ is a lower bound of the 
original GS energy \cite{dual}, i.e.

\begin{eqnarray}
&
e_{\infty}( J^U_\parallel(FA), J^U_\perp(FA)) \leq
e_\infty( J_\parallel(AA) , J_\perp(AA)) & 
\label{b8} \\
& 
e_\infty( J^U_\parallel(AF), J^U_\perp(AF)) \leq
e_\infty( J_\parallel(AF) , J_\perp(AF))  &  \nonumber
\end{eqnarray}

In fig. 5 we show the validity of these inequalities, which
in the strong coupling limit almost  become identities. 
In fig. 6 we show the correlation legths for both $AA$ and the
transformed $FA$ ladders. Again in the strong coupling limit they
become very close.

The $T$ transformation consists in the replacement of the
vertical bonds by diagonal ones, i.e.

\begin{eqnarray}
& J^T_\parallel  = J_\parallel &  \label{b9} \\
& J^T_\perp = J_\perp \; 
\langle {\bf S}_1(n) \cdot {\bf S}_1(n+1) \rangle/ 
\langle {\bf S}_1(n) \cdot {\bf S}_2(n+1) \rangle & \nonumber
\end{eqnarray}

\noindent which leads to the following action on magnetic
structures,

\begin{equation}
\begin{array}{ccc} 
AA & \stackrel{T}{\rightarrow} & AF \\
J_\parallel( AA) > 0 & {\rightarrow} & J_\parallel^T(AF) > 0 \\
J_\perp(AA) > 0 & {\rightarrow} &  J_\perp^T(AF) < 0  
\end{array}
\label{b10}
\end{equation}

\begin{equation}
\begin{array}{ccc} 
FA & \stackrel{T}{\rightarrow} & FA \\
J_\parallel( FA) < 0 & {\rightarrow} & J_\parallel^T(FA) <0 \\
J_\perp(FA) > 0 & {\rightarrow} &  J_\perp^T(FA) >0  
\end{array}
\label{b11}
\end{equation}

In fig.7 we plot the energies associated to the $FA$ ladder
and its $T$ transformed, which satisfies the inequality,

\begin{eqnarray}
&
e_{\infty}( J^T_\parallel(FA), J^T_\perp(FA)) \leq
e_\infty( J_\parallel(FA) , J_\perp(FA)) & 
\label{b12} 
\end{eqnarray}

The convergence of both curves in the
weak coupling is in agreement with the bosonization arguments 
employed in \cite{dual}.

Finally the $S$ transformation is defined by the replacement of
vertical bonds by horizontal ones and viceversa,

\begin{eqnarray}
&  J^S_\parallel  = \frac{1}{2} \; J_\perp \;
\langle {\bf S}_1(n) \cdot {\bf S}_2(n) \rangle/ 
\langle {\bf S}_1(n) \cdot {\bf S}_1(n+1) \rangle
& \label{b13} \\
& J^S_\perp =  2 \; J_\parallel \; 
\langle {\bf S}_1(n) \cdot {\bf S}_1(n+1) \rangle/ 
\langle {\bf S}_1(n) \cdot {\bf S}_2(n) \rangle & \nonumber 
\end{eqnarray}

\noindent 
The factors 2 and 1/2 are explained by the fact that 
there are two leg-bonds for each rung-bond. 
Eqs.(\ref{b13}) imply,

\begin{equation}
\begin{array}{ccc} 
AF & \stackrel{S}{\rightarrow} & FA \\
J_\parallel( AF) > 0 & {\rightarrow} & J_\parallel^S(FA) <0 \\
J_\perp(AF) < 0 & {\rightarrow} &  J_\perp^S(FA) >0  
\end{array}
\label{b14}
\end{equation}

\begin{equation}
\begin{array}{ccc} 
AA & \stackrel{S}{\rightarrow} & AA \\
J_\parallel( AA) > 0 & {\rightarrow} & J_\parallel^S(AA) >0 \\
J_\perp(AA) > 0 & {\rightarrow} &  J_\perp^S(AA) >0  
\end{array}
\label{b15}
\end{equation}

In fig. 8 we plot the energies of the $AA$ ladder and its transformed 
which satisfy the inequality,

\begin{eqnarray}
& e_{\infty}( J^S_\parallel(AA), J^S_\perp(AA)) \leq
e_\infty( J_\parallel(AA) , J_\perp(AA)) & 
\label{b16} 
\end{eqnarray}

Note  that in the region $J_\parallel \sim J_\perp$ 
both energies get very closed.
Fig. 9 shows the spin correlation length for the AA ladder
and its $S$ transformed, displaying the same pattern as fig. 8. 
In summary we have found  further numerical evidence
of the duality properties of the 2-leg ladder
proposed in \cite{dual}.

\subsection*{$AA_1$-ladder:
Short-range string order}

In table 5 we show the GS energy density and the spin-correlation
length of the ladder $AA_1$. Observe that the correlation length
is longer that the one of the spin 1/2  ladder.

\begin{center}
\begin{tabular}{ccc}
\hline
$J_\parallel/J_\perp$ & $-e_\infty/2 J_\perp$ & $\xi$ \\ 
\hline
\hline
0.0 & $1.000000$ & .00000  \\ 
\hline
$0.2$ & 1.055719  & 1.0114 \\ 
\hline
$0.4$ & 1.206557 & 1.8318 \\
\hline
$0.6$ & 1.407358 & 2.3852 \\
\hline
$0.8$ & 1.631166 & 2.6762 \\
\hline
$1.0$ & 1.867327 & 2.8227 \\
\hline
$1.25$ & 2.172905 & 2.9042  \\
\hline
$1.66$ & 2.688880 & 2.9286 \\
\hline
\end{tabular}
\end{center}
\begin{center}
Table 5. GS energy per site  and correlation length of the  $AA_1$
ladder.
\end{center}

As mentioned in the introduction 
a spin 1 chain has a long range topological order (LRTO)
characterized
by a non-vanishing $g(\infty)$. In appendix C we give 
an  analytical expression for  $g(\infty)$ in terms of the MP 
parameters of the spin 1 chain.

However when two spin 1 chains are 
coupled antiferromagnetically the LRTO 
disappears and the string order parameter 
$g(\ell)$ decays exponentially as $e^{- \ell/\xi^{\rm st}}$.
We call $\xi^{\rm st}$ the string correlation length, and
its value together with the spin correlation length are shown
in fig. 10 as functions of the ratio $J_\parallel/J_\perp$.
In the weak  coupling limit where 
$J_\parallel/J_\perp \rightarrow \infty $ we expect 
$\xi^{\rm st}$ to diverge, recovering in that way the
LRTO of the uncoupled chains.  The value of $\xi^{\rm st}$ 
is obtained by the formula (\ref{19}) with $x_p$ the highest
eigenvalue of the operator 
$\widehat{ e^{i \pi S^z_1} }$ (see appendix C).

An intuitive way to understand the breaking
of the LRTO is given by the AKLT picture of ref \cite{AKLT}.
An AKLT state is a valence bond state where every spin 1
is represented as a symmetrized product of two spins 1/2,
and such that every of these ``elementary'' spins is linked 
by a bond to one of the spins 1/2  on its neighbours.  
In this way all the spins of the chain are connected 
by a sucession of nearest neighbour links. When we
couple antiferromagnetically two spin 1 chains 
there is the possibility that two parallel 
bonds along the legs 
become two  parallel bonds along the rungs as shown in fig. 11.
Thus  the two infinite parallel arrays of connected bonds,
characteristic of the uncoupled chains, effectively breaks
into a collection of fluctuating
islands whose size is of the order of 
$\xi^{\rm st}$. Everyone  of these islands is a sort
of closed spin 1 chain ( fig.11).

The finite value of $\xi^{\rm st}$ at the origin of fig.10 
is due to the fact that $e^{i \pi S^z_1}$ has indeed a finite
value when computed  on the singlet formed by two spins 1
on a rung,

\begin{equation}
\langle e^{i \pi S^z_1} \rangle_{\rm rung} =
\sum_{m = \pm 1, 0} \; (-1)^m \langle 1 m 1 -m| 0 0  \rangle^2
= -  \frac{1}{3} 
\label{b17}
\end{equation}

\noindent which leads to
$\xi^{\rm st}(J_\parallel=0) = 1/{\rm ln}3$. 
Fig.10 suggests the existence of 3 different regimes.
In the weak coupling regime where $\xi^{\rm st} > \xi$ 
the ladder can be effectively considered as a collection
of weakly interacting closed spin 1 chains. In the strong
coupling regime where  $\xi^{\rm st} > \xi$ the bonds are meanly
along the rungs and the interbond coupling is small. Finally
there is an intermediate region, with  $\xi^{\rm st} < \xi$,  
where the islands of spins interact strongly with their
neighbours.

\subsection*{The  $AA_{3/2}$-ladder}
 
In table 6 we give the GS energy densities  and
spin correlation lengths  of the
ladder $AA_{3/2}$. As one may expect the correlation length
is longer than for the spin 1 and 1/2 ladders. 
This fact  agrees with the results obtained by mapping the 
spin  ladders into the NLSM \cite{Sen,Sie,M}.

\begin{center}
\begin{tabular}{ccc}
\hline
$J_\parallel/J_\perp$ & $-e_\infty/2 J_\perp$ & $\xi$ \\ 
\hline
\hline
0.0 & $1.875000$ & 0.0000  \\ 
\hline
$0.2$ & 2.054760   & 1.8760 \\ 
\hline
$0.4$ & 2.449827 & 3.3099 \\
\hline
$0.6$ & 2.911353 & 3.9475 \\
\hline
$0.8 $ & 3.400562 & 4.2401 \\
\hline
$1.0$ & 3.904988 & 4.3829 \\
\hline
$1.25$ & 4.548607 & 4.4624  \\
\hline
$1.66$ & 5.623131 & 4.4900 \\
\hline
\end{tabular}
\end{center}
\begin{center}
Table 6. GS energy per site   and correlation length of the 
$AA_{3/2}$ ladder.
\end{center}

\subsection*{AKLT states for ladders}

The spin 3/2 2-ladder offers the possibility of constructing
an AKLT state with a valence bond connecting 
every spin 3/2 to its three nearest neighbours. More generally,
let us consider a ladder with spin $S \geq 3/2$ and three
integers $p, q, r \geq 1$ satisfying the eq. $2 S = p + q + r$.
Then  one can define an AKLT state, denoted by the triplet 
$(p,q,r)$, by linking  the $2S$ ``elementary spinors''
of each spin to the ones in 
its neighbours following the pattern shown
in fig.(12). The AKLT states $(p,q,r)$ and $(q,p,r)$ 
when $p \neq q$  correspond to dimerized ladders
and they differ by the translation of one unit space
along the legs.

The  spin 3/2 AKLT  ladder corresponds
in the above notation to $(1,1,1)$. This state contains in fact
a spin 0 and a spin 1 state which  can be generated by the MP equation
(\ref{a1}) where the amplitudes $A_{J_1 J_2}^\lambda$ are
given by 9-j symbols,

\begin{equation}
A_{J_1 J_2}^\lambda= 3 \; \sqrt{ (2 J_2 +1) (2 \lambda +1)}
\; \left\{ 
\begin{array}{ccc} {1}/{2} & {1}/{2} & J_2 \\
                    {3}/{2} &  {3}/{2} & \lambda \\
                   1  & 1 & J_1 \end{array}
\right\}
\label{b18}
\end{equation}

In this eq. $J_1, J_2 = 0 $ and 1, while $\lambda = 0,1,2$. 

The proof of (\ref{b18}) follows from the definition
of the 9-j symbols 
as the coefficients that give 
the change of basis when coupling in two different  ways
4 angular momenta, namely \cite{group}

\begin{equation}
\psi\left( j_1 j_3 (J_{13}) j_2 j_4 (J_{24}) J \right)
\label{b19}
\end{equation}

\[
= \sum_{J_{12} J_{34} } \; \sqrt{ (2 J_{12} + 1 )
(2 J_{34} + 1 )
(2 J_{13} + 1 )(2 J_{24} + 1 )} 
\]

\[
\times 
\left\{ 
\begin{array}{ccc} j_1 & j_2 & J_{12} \\
                   j_3 & j_4 & J_{34} \\
                   J_{13}  & J_{24} & J \end{array}
\right\} \; \psi\left( j_1 j_2 (J_{12}) j_3 j_4 (J_{34}) J \right)
\]

\noindent
where $\psi\left( j_1 j_3 (J_{13}) j_2 j_4 (J_{24}) J \right)$ is 
a state with angular momentum $J$ obtained by the tensor
product decomposition $J_{13} \otimes J_{24} 
\rightarrow J$, 
which in turn are obtained by 
the decompositions $j_1 \otimes j_3 \rightarrow J_{13}$ and
$j_2 \otimes j_4 \rightarrow J_{24}$.

One may check that the normalization conditions (\ref{a5}) holds
for (\ref{b18}), as a consequence of the orthogonality conditions
satisfied by the 9-j symbols \cite{group}. 
The GS energy per site and the spin correlation length 
of the AKLT state (\ref{b18}) in the case where 
$J_\parallel = J_\perp =J $ are given by,

\begin{equation}
e_\infty^{\rm AKLT}/2 J= - 3.263536, \;\; 
\xi^{\rm AKLT}= 1.116221 
\label{AKLT}
\end{equation}

This state has a much shorter correlation length
than the MP state that minimizes the GS energy of the
$AA_{3/2}$-ladder (see table 6). The GS energies
of both states are also quite different.
We conclude from these facts  that 
the spin 3/2 AKLT state does not give
a good description 
of the GS of the $AA_{3/2}$-ladder.

A generic AKLT state of the type $(p,q,r)$ when $p \neq q$ 
has to be described by alternating  MP amplitudes depending
on the eveness of the site. Thus for even sites one has

\begin{equation}
A_{J_1 J_2}^\lambda= (q +r+1) \; \sqrt{ (2 J_2 +1) (2 \lambda +1)}
\; \left\{ 
\begin{array}{ccc} \frac{p}{2} & \frac{p}{2} & J_2 \\
                   S &  S & \lambda \\
                   \frac{q+r}{2}  & \frac{q+r}{2} & J_1 \end{array}
\right\}
\label{b20}
\end{equation}

\noindent where $J_1 = 0, \dots, q ; J_2 =0, \dots , p$
and $\lambda =0, \dots, 2S-r$. 
For odd sites the corresponding MP amplitudes
are obtained by interchanging  $p$ and $q$ in (\ref{b20}).

\section*{V) Conclusions and Prospects}

Let us summarize 
the main results obtained in this  paper.

\begin{itemize}

\item We have presented  a rotational invariant formulation of the 
MPM which allow us to express the GS energy density,
the correlation length and  the string order parameter,
in terms of invariant objects. This reduces
considerably the number of independent MP parameters
used in the minimization process.

\item We have improved the numerical results 
concerning the GS energy density and spin correlation 
length obtained
previously with other approximate methods as those of
references \cite{SMA,mean}. The  consideration of MP ansatzs 
with multiple states per spin 
will certainly lead to better results.

\item We have shown the equivalence between the ladder $AF_{1/2}$ 
and the spin 1 antiferromagnetic Heisenberg chain. 
The MPM applied to both systems shows strong
numerical coincidences for the GS energy and 
correlation length. This agrees with the results obtained
previously by other methods \cite{Hi,Wh2,Wa}

\item We have found numerical evidences for the duality properties
proposed in \cite{dual}
for  the spin ladders with magnetic structures
$AA, AF$ and $FA$.

\item We have shown that there is a  breaking of the long range topological
order of the spin 1 chains when they are coupled  in  
a 2 legged ladder. A physical picture of the GS of 
the spin 1 ladder is given in terms of 
resonating closed spin 1 chains.

\item We have  constructed  AKLT states for 2 legged 
ladders with spin $S \geq 3/2$,
showing that the corresponding 
MP parameters are given by 9-j
symbols.

\item We have suggested  a relation between the MPM and the DMRG
based on the density matrix that appear in both methods
(see also \cite{D}). We conjecture that 
the  minimization of the GS energy $e_\infty$
can be transformed 
into an eigenvalue problem on a superblock $B_N  \bullet B^R_N$.

\end{itemize}

In summary we have shown 
the adequacy of the MPM to study the 2 legged
ladder, specially in 
the strong and intermediate coupling regimes.
This is made possible from the 
fact that these ladders are 
finitely correlated. Hence one may expect that
even spin ladders with a finite number of legs 
could  be described by the same technique,
although with a larger number of states  $m$.  
On the other hand odd legged ladders are not finitely
correlated and they cannot be properly  described 
in the large $N$ limit within the 
actual formulation of the MPM. 
An interesting problem is the application of the MPM
to 2D systems, which can be thougth of as 
ladders with a large number of legs. 
It is clear that one should choose a 
collection of the most representative
states for the rungs to be added after each iteration
of the MP recurrence equation.

{\bf Acknowledgements}: JMR acknowledges financial 
support from a Basque Government
FPI grant as well as from CICYT, contract AEN95-0590, and CIRIT, 
contract GRQ93-1047. 
Also he thanks G. Sierra, 
J. Dukelsky and M.A. Mart{\'\i}n-Delgado for hospitality at the
IMAFF (CSIC), Madrid, where this work was performed.
GS and MAMD acknowledges support from
the DIGICYT under contract No. PB96/0906 and 
JD acknowledges support from the DIGICYT under
contract No. PB95/0123.

\section*{Appendix A: The MP ansatz and the Grassmannian manifolds}

In this appendix we 
shall give a proof of eq.(\ref{4}) which
gives a precise mathematical meaning of the coefficients
$A_{\alpha \beta}[s]$ defining  a generic MP ansatz.

In the  r.h.s. of eq.(\ref{1})
we have a generic vector of dimension $n = m m^*$ while
on its l.h.s. the vector has dimension $m$. Hence
eq.(\ref{1}) amounts to a choice of a $m$-dimensional linear
subspace of ${\bf R}^n$ in the case of 
$A_{\alpha \beta}[s]$ real
or a complex subspace of 
${\bf C}^n$ in the case of
$A_{\alpha \beta}[s]$ complex. Let us call the
set of all these subspaces as 
$M_{n,m}({\bf R})$ and $M_{n,m}({\bf C})$ for 
$A_{\alpha \beta}[s]$ real and
complex respectively. 
The group $O(n)$ ( resp.  $U(n)$ ) 
acts transitively on $M_{n,m}({\bf R})$ ( resp. $M_{n,m}({\bf C})$),
which leads  to the result \cite{enciclo}

\begin{equation}
M_{n,m}({\bf R})= \frac{O(n)}{ O(m) \otimes O(n-m)} 
\label{A1}
\end{equation}

\begin{equation}
M_{n,m}({\bf C})= \frac{U(n)}{ U(m) \otimes U(n-m)} 
\label{A2} 
\end{equation}

In (\ref{A1}) the groups $O(m)$ and $O(n-m)$ are
identified with the subgroups  of $O(n)$ consisting
of those elements leaving fixed every vector
of a given $(n-m)$-dimensional subspace and of its 
orthogonal complement, respectively. Similar
arguments lead to eq.(\ref{A2}).
$M_{n,m}({\bf R}) (  M_{n,m}({\bf C})$) are called the
real (complex) Grassmannian manifolds.
Taking $n=m m^*$ in (\ref{A1}) we get eq.(\ref{4}).

As a simple illustration of these eqs. 
let us consider the case of a MP ansatz that 
generates  a single state $|GS\rangle_N$  ($m=1$), i.e.

\begin{equation}
|GS \rangle_{N} = \sum_{s} A[s] \;
|s\rangle_{N} \otimes
|GS \rangle_{N-1} \;  
\label{A3}
\end{equation}

\noindent with  $A[s] \in {\bf R}$. 
The normalization  condition (\ref{2}) reads,

\begin{equation}
\sum_{s=1}^{m^*} \; A[s]^2 = 1
\label{A4}
\end{equation}

Thus $A[s]$ 
belongs to the $(m^*-1)$-dimensional sphere
$SO(m^*)/SO(m^*-1)$. 
Upon the identification of $A[s]$
and $-A[s]$ we get the $(m^*-1)$-real proyective space 
$M_{m^*,1}({\bf R}) = SO(m^*)/SO(m^*-1) \otimes {\bf Z}_2$.

\section*{Appendix B: The  Rotational Invariant MPM  }

{\bf Group theoretical preliminaries}

Before we give the proof of the main formulas of section III
we shall review some basic definitions and  results in group 
theory \cite{group}.

An irreducible tensor with angular momentum $k$  
is an operator ${T}^{(k)}_M \; ( M = k, \cdots, -k)$
which satisfies the following commutation relations with the
total angular momentum operator ${\bf J}$,

\begin{eqnarray}
& [ J_z , {T}^{(k)}_M ] = M \; {T}^{(k)}_M & \label{B1} \\
& [ J_x \pm i \; J_y  , {T}^{(k)}_M ] = 
\sqrt{k ( k+1) - M ( M \pm 1) } \; 
{T}^{(k)}_{M \pm 1}  & \nonumber 
\end{eqnarray}

The scalar product of two irreducible tensors ${\bf T}^{(k)}$
and   ${\bf U}^{(k)}$ with the same spin $k$ is defined by,

\begin{equation}
{\bf T}^{(k)} \cdot {\bf U}^{(k)} =
\sum_{M= -k}^k (-1)^{-M} \; T^{(k)}_M \;  U^{(k)}_{-M}
\label{B2}
\end{equation}

The Wigner-Eckart theorem reads,

\begin{eqnarray}
& \langle JM | T^{(k)}_\mu | J' M' \rangle & \label{B3} \\
& =  (-1)^{J-M} \; \left( \begin{array}{ccc} J & k & J' \\
                                          -M & \mu & M' \end{array}
\right) \; ( J || {\bf T}^{(k)} || J' )  & \nonumber 
\end{eqnarray}

\noindent
where the 3-j symbol is related to the CG coefficient by

\begin{equation}
\left( \begin{array}{ccc} J & k & J' \\
                                          -M & \mu & M' \end{array}
\right) = \frac{ (-1)^{J-k-M'} }{\sqrt{2 J' +1} } \;
\langle J -M k \mu | J' - M' \rangle 
\label{B4}
\end{equation}

The quantity $( J || {\bf T}^{(k)} || J' ) $ in (\ref{B3}) 
is called the reduced matrix 
element of the operator $ {\bf T}^{(k)}$ . As an example we give
the reduced matrix element of the spin operator ${\bf S}$,

\begin{equation}
( S || {\bf S} || S ) = \sqrt{ S ( S+1) (2 S +1)} 
\label{B5}
\end{equation}

Let $|\alpha_1 j_1 \alpha_2 j_2 J M \rangle$
be a state with total angular momenta $J$ and third component
$M$,  appearing  in the tensor product decomposition
$(\alpha_1 j_1) \otimes  (\alpha_2 j_2)$, where $(\alpha j)$
denotes a state with total angular momentum $j$ and $\alpha$
labels other possible quantum numbers. 
We shall need below the following results.

\begin{eqnarray}
& \langle \alpha_1 j_1 \alpha_2 j_2 J M| ( {\bf T}^{(k)}_1 
\cdot {\bf T}^{(k)}_2 )
|\alpha'_1 j'_1 \alpha'_2 j'_2 J' M' \rangle  & \nonumber \\
&= \delta_{J J'} \; \delta_{M M'} \; (-1)^{j_2 + J + j'_1} 
\; \left\{ \begin{array}{ccc} j_1 & j_2 & J \\
                         j'_2 & j'_1 & k \end{array} \right\}
& \label{B6} \\
& \times (\alpha_1 j_1 || {\bf T}^{(k)}_1 || \alpha'_1 j'_1 )\;
(\alpha_2 j_2 || {\bf T}^{(k)}_2 || \alpha'_2 j'_2 ) & \nonumber 
\end{eqnarray}

\begin{eqnarray}
& ( \alpha_1 j_1 \alpha_2 j_2 J || {\bf T}^{(k)}_1 
||\alpha'_1 j'_1 \alpha'_2 j'_2 J' )  & \nonumber \\
&= \delta_{\alpha_2 \alpha'_2} \delta_{j_2 j'_2} \;  \; 
(-1)^{j_1 + j_2 + J' +k} 
\; \left\{ \begin{array}{ccc} j_1 & J & j_2 \\
                         J' & j'_1 & k \end{array} \right\}
& \label{B7} \\
& \times  \sqrt{(2 J +1) (2 J' +1)} \;
(\alpha_1 j_1 || {\bf T}^{(k)}_1 || \alpha'_1 j'_1 )   & \nonumber 
\end{eqnarray}

\begin{eqnarray}
& ( \alpha_1 j_1 \alpha_2 j_2 J || {\bf T}^{(k)}_2 
||\alpha'_1 j'_1 \alpha'_2 j'_2 J' )   & \nonumber \\
&= \delta_{\alpha_1 \alpha'_1} \delta_{j_1 j'_1} \;  \; 
(-1)^{j_1 + j'_2 + J +k} 
\; \left\{ \begin{array}{ccc} j_2 & J & j_1 \\
                         J' & j'_2 & k \end{array} \right\}
& \label{B8} \\
& \times \sqrt{(2 J +1) (2 J' +1)} \;
(\alpha_2 j_2 || {\bf T}^{(k)}_2 || \alpha'_2 j'_2 )\;
& \nonumber 
\end{eqnarray}

The subindices 1 and 2 in ${\bf T}^{(k)}_1$ and   ${\bf T}^{(k)}_2$
mean that the corresponding operators acts on the states
labelled as $(\alpha_1 j_1)$ and 
$(\alpha_2 j_2)$ respectively.

{\bf Recursion relations for the scalar product of invariant
tensors}

We want to prove eq.(\ref{a7}).

Using eq.(\ref{a1}) we easily get for $N > n > m$,

\begin{equation}
_N\langle J_1 M| \;
{\cal O}^{(k,A)}(n) \cdot {\cal O}^{(k,B)}(m)
\; | J_1 M \rangle_N 
\label{B9} 
\end{equation}

\[=
\sum_{J_2} \; 
T_{J_1, J_2} \; _{N-1}\langle J_2 M| \;
{\cal O}^{(k,A)}(n) \cdot {\cal O}^{(k,B)}(m)
\; | J_2 M \rangle_{N-1} 
\]

\noindent where $T_{J_1, J_2}$ is given in (\ref{a8}).
Iterating (\ref{B9}) $N-n$ times we reach the situation where
$N=n$. This produces  the term $T_{J_1, J_2}^{N-n}$ in (\ref{a7}).
Next we need to compute the matrix element,

\begin{equation}
_n\langle J_1 M| \;
{\cal O}^{(k,A)}(n) \cdot {\cal O}^{(k,B)}(m)
\; | J_1 M \rangle_n 
\label{B10} 
\end{equation}

\[=
\sum_{J_2 J_3 \lambda_2 \lambda_3} \; 
\left( A_{J_1 J_2}^{\lambda_2} \right)^*
\;  A_{J_1 J_3}^{\lambda_3} 
\]

\[ \times 
_n\langle (\lambda_2 J_2 ), J_1 M| \;
{\cal O}^{(k,A)}(n) \cdot {\cal O}^{(k,B)}(m)
\; | (\lambda_3 J_3), J_1 M \rangle_n 
\]

The matrix element on the r.h.s. of (\ref{B10})
has the form described in (\ref{B6}), which yields,

\begin{eqnarray}
& _n\langle (\lambda_2 J_2 ), J_1 M| \;
{\cal O}^{(k,A)}(n) \cdot {\cal O}^{(k,B)}(m)
\; | (\lambda_3 J_3), J_1 M \rangle_n 
& \label{B11} \\
&= (-1)^{J_1 + J_2 + \lambda_3} \; 
 \left\{ \begin{array}{ccc} \lambda_2 & J_2 & J_1 \\
                         J_3 & \lambda_3 & k \end{array} \right\}&
\nonumber \\
& \times \; 
_n(\lambda_2 || {\cal O}^{(k,A)}(n)|| \lambda_3)_n \; 
_{n-1}(J_2 || {\cal O}^{(k,B)}(m)|| J_3 )_{n-1} & \nonumber
\end{eqnarray}

Introducing (\ref{B11}) into (\ref{B10}) we find

\begin{equation}
_n\langle J_1 M| \;
{\cal O}^{(k,A)}(n) \cdot {\cal O}^{(k,B)}(m)
\; | J_1 M \rangle_n 
\label{B12} 
\end{equation}

\[=
\sum_{J_2 J_3 } \; \widehat{\cal O}^{(k,A)}_{J_1, J_2 J_3}
\; _{n-1}( J_2 ||  {\cal O}^{(k,B)}(m) || J_3)_{n-1} 
\]

\noindent where

\begin{eqnarray}
&\widehat{\cal O}^{(k,A)}_{J_1, J_2 J_3} = 
\sum_{\lambda_2, \lambda_3} \; 
\left( A_{J_1 J_2}^{\lambda_2} \right)^*
A_{J_1 J_3}^{\lambda_3} & \label{B13} \\
& \times (-1)^{\lambda_3 + J_1 + J_2} 
\; \left\{ \begin{array}{ccc} 
\lambda_2 &  J_2  & J_1 \\
J_3 &  \lambda_3  & k \end{array}
\right\} ( \lambda_2 ||{\cal O}^{(k,A)}||\lambda_3) &
\nonumber 
\end{eqnarray}

The next step is to apply the MP ansatz (\ref{a1}) to

\begin{equation}
_{n}( J_1 ||  {\cal O}^{(k,B)}(m) || J_2)_{n} 
\label{B14}
\end{equation}

\[= \sum_{\lambda_1 \lambda_2} \; 
\left( A_{J_1 J_3}^{\lambda_1} \right)^*
\;  A_{J_2 J_4}^{\lambda_2} 
\]

\[ \times 
_n( (\lambda_1 J_3 ), J_1 || \;
{\cal O}^{(k,B)}(m)
\; || (\lambda_2 J_4), J_2  )_n 
\]

For $n > m$ we can use (\ref{B8}), getting

\begin{equation}
_n( (\lambda_1 J_3 ), J_1 || \;
{\cal O}^{(k,B)}(m)
\; || (\lambda_2 J_4), J_2  )_n 
\label{B15}
\end{equation}

\[ = \delta_{\lambda_1 \lambda_2} \; (-1)^{ \lambda_1 + J_1 + J_4 +k}
\; \sqrt{(2 J_1 +1)( 2 J_2 +1)} 
\]

\[ \times 
\left\{ \begin{array}{ccc} J_3 & J_1 & \lambda_1 \\
                         J_2 & J_4 & k \end{array} \right\}
\; _{n-1}( J_3 || \; {\cal O}^{(k,B)}(m) \;  || J_4)_{n-1}
\]

Plugging (\ref{B15}) into (\ref{B14}) we get,

\begin{equation}
_{n}( J_1 ||  {\cal O}^{(k,B)}(m) || J_2)_{n} 
\label{B16}
\end{equation}

\[= \sum_{J_3 J_4}\; (T_k)_{J_1 J_2, J_3 J_4} \; 
_{n-1}( J_3 ||  {\cal O}^{(k,B)}(m) || J_4)_{n-1} , \;\; (n > m)
\]

\noindent where $(T_k)_{J_1 J_2, J_3 J_4}$ is defined in 
(\ref{a9}). The term $T^{n-m-1}_k$ in (\ref{a7}) results
from the iteration of (\ref{B16}) until one gets $n=m$.  
In the case when  $n=m$ in  (\ref{B15}) we should apply
(\ref{B7}) obtaining

\begin{equation}
_n( (\lambda_1 J_3 ), J_1 || \;
{\cal O}^{(k,B)}(n)
\; || (\lambda_2 J_4), J_2  )_n 
\label{B17}
\end{equation}

\[ = \delta_{J_3 J_4} 
\; (-1)^{ \lambda_1 + J_2 + J_3 +k}
\; \sqrt{(2 J_1 +1)( 2 J_2 +1)} 
\]

\[ \times 
\left\{ \begin{array}{ccc} \lambda_1 & J_1 & J_3 \\
                         J_2 & \lambda_2 & k \end{array} \right\}
\; _{n}( \lambda_1 || \; {\cal O}^{(k,B)}(n) \;  || \lambda_2)_{n}
\]

Introducing (\ref{B17}) into (\ref{B14}) we get,

\begin{equation}
_{n}( J_1 ||  {\cal O}^{(k,B)}(n) || J_2)_{n} = \sum_{J_3}
\; \widehat{\cal O}^{(k,B)}_{J_1 J_2,  J_3} 
\label{B18}
\end{equation}

\noindent where

\begin{eqnarray}
&\widehat{\cal O}^{(k,B)}_{J_1 J_2,  J_3} = 
\sum_{\lambda_1 \lambda_2} \; 
\left( A_{J_1 J_3}^{\lambda_1} \right)^*
A_{J_2 J_3}^{\lambda_2} (-1)^{\lambda_1 + J_2 + J_3 +k} & \label{B19}\\ 
&\times \sqrt{(2J_1+1)(2J_2+1)} 
\; \left\{ \begin{array}{ccc} 
\lambda_1 &  J_1  & J_3 \\
J_2 &  \lambda_2  & k \end{array}
\right\} ( \lambda_1 ||{\cal O}^{(k,B)}||\lambda_2) &
\nonumber 
\end{eqnarray}

This ends the proof of eq.(\ref{a7}).

{\bf Recursion relation of the energy expectation values}

We shall not give here the explicit
proof of eq.(\ref{a18}) since it is quite analogous to the
one performed  in the previous paragraph. We shall simply state
the result.

The matrix $\hat{h}_{J_1,J_2}$ appearing in (\ref{a18})
is given by the sum

\begin{equation}
\hat{h}_{J_1,J_2} = \hat{h}_{J_1,J_2}^{(1)} + 
\hat{h}_{J_1,J_2}^{(2)}
\label{B20}
\end{equation}

\noindent where

 \begin{equation}
\widehat{h}^{(1)}_{J_1, J_2} = J_{\perp}
\sum_{\lambda} \; \left( \frac{1}{2} \lambda (\lambda +1)
- S ( S+1) \right) |A^\lambda_{J_1 J_2}|^2 
\label{B21}
\end{equation}

\[
\widehat{h}^{(2)}_{J_1, J_4} = 2J_{\parallel} \;
\sum_{J_2 J_3 J_4, \lambda_1, \dots, \lambda_4} 
\; (A^{\lambda_1 }_{J_1 J_2}
A^{\lambda_3 }_{J_2 J_4} )^*
A^{\lambda_2}_{J_1 J_3}
A^{\lambda_4}_{J_3 J_4}
\nonumber 
\]

\begin{equation}
\times (-1)^{1+\lambda_3 + \lambda_4} \; 
\xi^{ \lambda_2 \lambda_1}_{J_2 J_3 J_1} \;
\xi^{ \lambda_3 \lambda_4}_{J_3 J_2 J_4}
\label{B22}
\end{equation}

\noindent and

\[
\xi^{ \lambda_1 \lambda_2}_{J_1 J_2 J_3}
=  (-1)^{J_1 + J_3} \;
\sqrt{(2J_1+1)(2 \lambda_1+1) (2 \lambda_2 +1)} 
\]

\[ \times 
\sqrt{S(S+1)(2S+1)}
\left\{ \begin{array}{ccc} 
\lambda_1 &  \lambda_2  & 1 \\
J_1 &  J_2  & J_3 \end{array}
\right\}
\left\{ \begin{array}{ccc} 
\lambda_1 &  \lambda_2  & 1 \\
S &  S  & S \end{array}
\right\}
\]

\noindent
where the following  property for the 6-j symbol
with an element equal 1 has been used \cite{group}:

\[
\left\{ \begin{array}{ccc} 
\lambda_1 &  \lambda_2  & 1 \\
J_1 &  J_2  & J_3 \end{array}
\right\} =
\left\{ \begin{array}{ccc} 
\lambda_2 &  \lambda_1  & 1 \\
J_2 &  J_1  & J_3 \end{array}
\right\}
\nonumber
\]

\section*{Appendix C: The string order
parameter of spin 1 chain and ladder}

Let us first consider the spin 1 chain. The MP ansatz
is given simply by,

\begin{equation}
|J_1 M_1\rangle_N = \sum_{J_2} \;
A_{J_1 J_2} \; |(1 J_2), J_1 M_1 \rangle_N 
\label{C1} 
\end{equation}

\noindent where the state
$|(1 J_2), J_1 M_1 \rangle_N$ reads as in (\ref{a2})
with $\lambda =1$. We shall choose half-integer
values of the angular momenta $J_1$ and $J_2$
which amounts to have a spin 1/2 at one end 
of the chain \cite{OR,D}.

We shall next show that the operators $T= \widehat{1}$
and $\widehat{ e^{ i \pi S^z} }$ have both an eigenvalue
equal to 1. Let us first of all write out explicitely
their components,

\begin{eqnarray}
&\left(T \right)_{J_1 M_1 J'_1 M'_1 ,J_2 M_2 J'_2 M'_2} & \nonumber \\
&= \delta_{M_1-M_2, M'_1-M'_2} \; A_{J_1 J_2} \; A_{J'_1 J'_2} &
\label{C2} \\
& \times \langle 1 M_1-M_2, J_2 M_2 | J_1 M_1 \rangle 
\langle 1 M'_1-M'_2, J'_2 M'_2 | J'_1 M'_1 \rangle & 
\nonumber
\end{eqnarray}

\begin{eqnarray}
&\left(\widehat{ 
e^{ i \pi S^z} } \right)_{J_1 M_1 J'_1 M'_1 ,J_2 M_2 J'_2 M'_2} &
\nonumber  \\
& = \delta_{M_1-M_2, M'_1-M'_2} \; (-1)^{M_1 - M_2}
\; A_{J_1 J_2} \; A_{J'_1 J'_2} &
\label{C3} \\
& \times \langle 1 M_1-M_2, J_2 M_2 | J_1 M_1 \rangle 
\langle 1 M'_1-M'_2, J'_2 M'_2 | J'_1 M'_1 \rangle & 
\nonumber 
\end{eqnarray}

The normalization conditions on $A_{J_1 J_2}$ read

\begin{equation}
\sum_{J_2} \; A^2_{J_1 J_2} = 1 , \,\, \forall \; J_1
\label{C4}
\end{equation}

Using these eqs. and the properties of the CG
coefficients,   
one can  verify that $v$ and $v^{\rm st}$ defined as

\begin{eqnarray}
& v_{J_1 M_1 J'_1 M'_1} = \delta_{J_1 J'_1} \; 
\delta_{M_1 M'_1} & \label{C5} \\
& v_{J_1 M_1 J'_1 M'_1}^{\rm st} = \delta_{J_1 J'_1} \; 
\delta_{M_1 M'_1} (-1)^{M_1 - 1/2}
& \nonumber 
\end{eqnarray}

\noindent
are right eigenvectors with eigenvalue 1 of 
the matrices $T$ and 
$\widehat{ e^{ i \pi S^z} }$ respectively.

Similarly the left eigenvectors associated to this
eigenvalue are given by,

\begin{eqnarray}
& \rho_{J_1 M_1 J'_1 M'_1} = \delta_{J_1 J'_1} \; 
\delta_{M_1 M'_1} \; \rho_{J_1}/(2 J_1 + 1)
 & \label{C6} \\
& \rho_{J_1 M_1 J'_1 M'_1}^{\rm st} = \delta_{J_1 J'_1} \; 
\delta_{M_1 M'_1} (-1)^{M_1 - 1/2}  \;  \rho_{J_1}/(2 J_1 + 1)
& \nonumber  
\end{eqnarray}

\noindent 
where $\rho_{J}$ is the left eigenvector with eigenvalue 1
of the matrix 
$T_{J_1 J_2} = A_{J_1 J_2}^2 $.

According to eq. (\ref{22}) the string order parameter 
$g(\infty)$ is given by the product
of two matrix elements which we compute below.

Let us first consider,

\begin{eqnarray}
& \langle \rho| \widehat{S^z} 
| v^{\rm st} \rangle & \label{C7} \\
&= \sum \rho_{J_1 M_1 J'_1 M'_1} \; 
\widehat{S^z}_{J_1 M_1 J'_1 M'_1, J_2 M_2 J'_2 M'_2} \; 
v_{J_2 M_2 J'_2 M'_2}^{\rm st}  & \nonumber 
\end{eqnarray}

The hated version of ${S^z}$ is given by,

\begin{eqnarray}
&\left(\widehat{S^z}
\right)_{J_1 M_1 J'_1 M'_1 ,J_2 M_2 J'_2 M'_2}&  \nonumber \\
& =\delta_{M_1-M_2, M'_1-M'_2} \; (M_1 - M_2)
\; A_{J_1 J_2} \; A_{J'_1 J'_2} &
\label{C8} \\
& \times  \langle 1 M_1-M_2, J_2 M_2 | J_1 M_1 \rangle 
\langle 1 M'_1-M'_2, J'_2 M'_2 | J'_1 M'_1 \rangle & 
\nonumber 
\end{eqnarray}

\noindent which together with (\ref{C6},\ref{C7}) 
lead to,

\begin{eqnarray}
& \langle \rho| \widehat{S^z}  
| v^{\rm st} \rangle = \sum \frac{\rho_{J_1}}{ 2 J_1 + 1} A_{J_1 J_2}^2 &
\label{C9} \\
& \times \; (-1)^{M_2 - 1/2} \; (M_1 - M_2) 
 \left( \langle 1 M_1 - M_2,
J_2 M_2 | J_1 M_1 \rangle \right)^2 & \nonumber 
\end{eqnarray}

Similarly we get

\begin{eqnarray}
& \langle \rho^{\rm st}| \widehat{S^z}  
| v \rangle = \sum \frac{\rho_{J_1}}{ 2 J_1 + 1} A_{J_1 J_2}^2 &
\label{C10} \\
& \times \; (-1)^{M_1 - 1/2} \; (M_1 - M_2) 
 \left( \langle 1 M_1 - M_2,
J_2 M_2 | J_1 M_1 \rangle \right)^2 & \nonumber 
\end{eqnarray}

Observing that

\begin{equation}
(-1)^{M_1 - 1/2} \; (M_1-M_2) = - 
(-1)^{M_2 - 1/2} \; (M_1-M_2)
\label{C11}
\end{equation}

\noindent
where $M_1 - M_2 =0, \pm 1$,  we obtain

\begin{equation}
\langle \rho^{\rm st}| \widehat{S^z}  
| v \rangle = - \langle \rho| \widehat{S^z}  
| v^{\rm st} \rangle 
\label{C12}
\end{equation}

\noindent which in turn implies

\begin{equation}
g(\infty) = -
\left( \langle \rho| \widehat{S^z} 
| v^{\rm st} \rangle \right)^2
\label{C13}
\end{equation}

Let us come back to eq.(\ref{C9}), which  can be written as

\begin{equation}
 \sum \frac{ - \rho_{J_1}}{ 2 J_1 + 1} A_{J_1 J_2}^2 
\; (-1)^{M_1 - 1/2} \; 
\langle (1 J_2) J_1  M_1| S^z_1 |
 (1 J_2) J_1 M_1 \rangle 
\label{C14}
\end{equation}

\noindent where  $S^z_1$ denotes the spin operator
acting on the spin 1.  Using the Wigner-Eckart theorem 
we get,

\begin{eqnarray}
& \langle (1 J_2) J_1  M_1| S^z_1 |
 (1 J_2) J_1 M_1 \rangle  & \nonumber \\ 
& = (-1)^{J_1 - M_1} \, 
\left( \begin{array}{ccc} J_1 & 1 & J_1 \\ - M_1 & 0 & M_1
\end{array} \right) \; ((1 J_2) J_1 || {\bf S}_1 ||(1 J_2) J_1)&
\label{C15} \\
& = \frac{M_1}{ \sqrt{ J_1( 2 J_1 +1)( J_1 +1)} }
\;  ((1 J_2) J_1 || {\bf S}_1 ||(1 J_2) J_1)& \nonumber 
\end{eqnarray}

The reduced matrix element appearing in (\ref{C15})
can be computed using (\ref{B7}),

\begin{eqnarray}
&((1 J_2) J_1 || {\bf S}_1 ||(1 J_2) J_1) & \nonumber \\
& =\sqrt{6} \; (-1)^{J_1 + J_2}\;  (2 J_1 + 1) 
\left\{ \begin{array}{ccc} 1 & J_1 & J_2 \\J_1 & 1 & 1 
\end{array} \right\} & \label{C16} \\ 
& = \frac{ \sqrt{2J_1 +1} (2 + J_1 ( J_1 +1) - J_2 ( J_2 + 1)) }{
2 \; \sqrt{ J_1 (J_1 +1 )}} & \nonumber
\end{eqnarray}

Substituting (\ref{C15},\ref{C16}) into (\ref{C14}) and performing the
sum over $M_1$ with the aid of the formula,

\begin{equation}
\sum_{M = -J}^J M (-1)^{M-\frac{1}{2}} = \; ( J + \frac{1}{2}) \; (-1)^{J-1/2},
\; (J : {\rm half} \;{\rm integer})
\label{C17}
\end{equation}

\noindent 
we get finally,

\[
\langle \rho| \widehat{S^z} 
| v^{\rm st} \rangle = \frac{1}{4} \sum \rho_{J_1} \, A^2_{J_1 J_2} 
\; (-1)^{J_1 - 1/2} 
\]

\begin{equation}
\times \frac{ 2 + J_1 ( J_1 +1) - J_2 ( J_2 +1)}{
J_1 ( J_1 +1) } 
\label{C18}
\end{equation}

 From eqs(\ref{C13},\ref{C18}) we immediately get the value
of $g(\infty)$ in
the AKLT case,

\begin{equation} 
{\rm AKLT:}  \; A_{\frac{1}{2} \frac{1}{2}}= 1 \; \rightarrow \; 
g(\infty) = - (2/3)^2
\label{C19}
\end{equation}

In ref \cite{D} the spin 1 Heisenberg chain was studied
with a MP ansatz built up with two states with 
$J= 1/2$ and $3/2$. The values of the MP parameters 
obtained in  \cite{D} are reproduced below

\begin{eqnarray}
& A_{\frac{1}{2}  \frac{1}{2} } =0.988995  , \; 
A_{ \frac{1}{2} \frac{3}{2} } =0.14795 & \nonumber \\
& A_{ \frac{3}{2} \frac{1}{2} } = -0.952887 , 
\; A_{ \frac{3}{2} \frac{3}{2} } = -0.303325 & \label{C20} \\
&\rho_{\frac{1}{2}} = 0.97646, \rho_{ \frac{3}{2} } = 0.023539 &
\nonumber   
\end{eqnarray}

Introducing (\ref{C20}) into eqs.(\ref{C13}) and (\ref{C18})
we get $g(\infty) = - 0.387$, which can be compared with 
the exact value given by -0.374325 \cite{WH}.
In \cite{OR} the spin 1 chain was studied
with a MP ansatz with two  spin 1/2
and two spin 3/2 states, which yields
$g(\infty) = -0.3759$. This shows again that  
MP ansatzs with multiplicity 
improve considerably the accuracy of the 
numerical results
\cite{OR,D}.

Let us go now to the spin 1 ladder. In section IV we gave an
intuitive argument which suggested that  the LRTO  of the single
spin 1 chains  is destroyed by the interchain coupling.
Next we show that this is indeed what happens.

Let us first write eqs (\ref{C2},\ref{C3}) in the case of ladders.

\begin{eqnarray}
&\left(T \right)_{J_1 M_1 J'_1 M'_1 ,J_2 M_2 J'_2 M'_2} & \nonumber \\
&= \delta_{M_1-M_2, M'_1-M'_2} 
\sum_{\lambda} \; A_{J_1 J_2}^\lambda \; A_{J'_1 J'_2}^\lambda &
\label{C21} \\
& \times \langle \lambda M_1-M_2, J_2 M_2 | J_1 M_1 \rangle 
\langle \lambda M'_1-M'_2, J'_2 M'_2 | J'_1 M'_1 \rangle & 
\nonumber
\end{eqnarray}

\begin{eqnarray}
&\left(\widehat{ 
e^{ i \pi S^z_1} } \right)_{J_1 M_1 J'_1 M'_1 ,J_2 M_2 J'_2 M'_2} &
\nonumber  \\
& = \delta_{M_1-M_2, M'_1-M'_2} \; \sum_{\lambda \lambda'} 
\; A_{J_1 J_2}^\lambda \; A_{J'_1 J'_2}^{\lambda'} &
\label{C22} \\
& \times \langle 1 M_1-M_2, J_2 M_2 | J_1 M_1 \rangle 
\langle 1 M'_1-M'_2, J'_2 M'_2 | J'_1 M'_1 \rangle & 
\nonumber \\
& \times \; \langle \lambda M_1-M_2 | e^{i \pi S^z_1} | \lambda' M_1 M_2 
\rangle & \nonumber 
\end{eqnarray}

\noindent where ${ S^z_1}$ denotes the spin operator acting
on the first leg of the ladder.
The vector $v_{J_1 M_1 J_1' M_1'}$ given  in (\ref{C5})
is an eigenvector with eigenvalue 1 
of the matrix $T$ defined  by  (\ref{C21}). This property
is a consequence of the normalization condition (\ref{a5}).
For the spin 1 chain the latter condition also guarantees
the existence of an eigenvalue 1 of the operator (\ref{C3}).
However this is not generally 
the case for the operator (\ref{C22}).

The last matrix element in (\ref{C22}) can be deduced
expressing the state $|\lambda \mu\rangle$ of the rung
in terms of the spin 1 states of every site,

\begin{equation}
|\lambda \mu \rangle = \sum_{m_1 m_2} \;
|1 m_1 \rangle_1 \; | s m_2 \rangle_2 \langle 1 m_1 1 m_2 | \lambda \mu 
\rangle  
\label{C23}
\end{equation}

We thus get

\begin{eqnarray}
&\left(\widehat{ 
e^{ i \pi S^z_1} } \right)_{J_1 M_1 J'_1 M'_1 ,J_2 M_2 J'_2 M'_2} &
\nonumber  \\
& = \delta_{M_1-M_2, M'_1-M'_2} \; \sum_{\lambda \lambda' m_1 m_2} 
\; A_{J_1 J_2}^\lambda \; A_{J'_1 J'_2}^{\lambda'} (-1)^{m_1} &
\label{C24} \\
& \times \langle 1 M_1-M_2, J_2 M_2 | J_1 M_1 \rangle 
\langle 1 M'_1-M'_2, J'_2 M'_2 | J'_1 M'_1 \rangle & 
\nonumber \\
& \times \; \langle 1 m_1 1 m_2 | \lambda M_1-M_2 \rangle 
 \langle 1 m_1 1 m_2 | \lambda' M_1-M_2 \rangle 
\rangle & \nonumber 
\end{eqnarray}

We can actually set up $M_1 = M_1'$ and $M_2 = M_2'$ in (\ref{C24})
since in the computation of the
string order parameter, 
the third component of the
angular momenta is preserved. We have computed the highest
eigenvalue $x_{\rm st}$ of the matrix (\ref{C24}), which turns out 
to be smaller 
than one. This shows that $g(\ell)$ decays exponentially
with a correlation length $\xi^{\rm st}$ whose value is obtained 
by the eq.

\begin{equation}
\xi^{\rm st} = -1/{\rm ln}|x_{\rm st}|
\label{C25}
\end{equation}


\section*{Figure Captions}

{\bf Fig.1} Graphical representation of the MP ansatz (\ref{a1})
in the case of the spin 1/2 ladder and basis $|J M\rangle_N$
with $J=0 $ and 1. Every  dot represents a  spin 1/2. 
A link between two dots 
denotes the formation of a singlet between the spins. Doted
lines denote symetrization of the spins encircled by them.

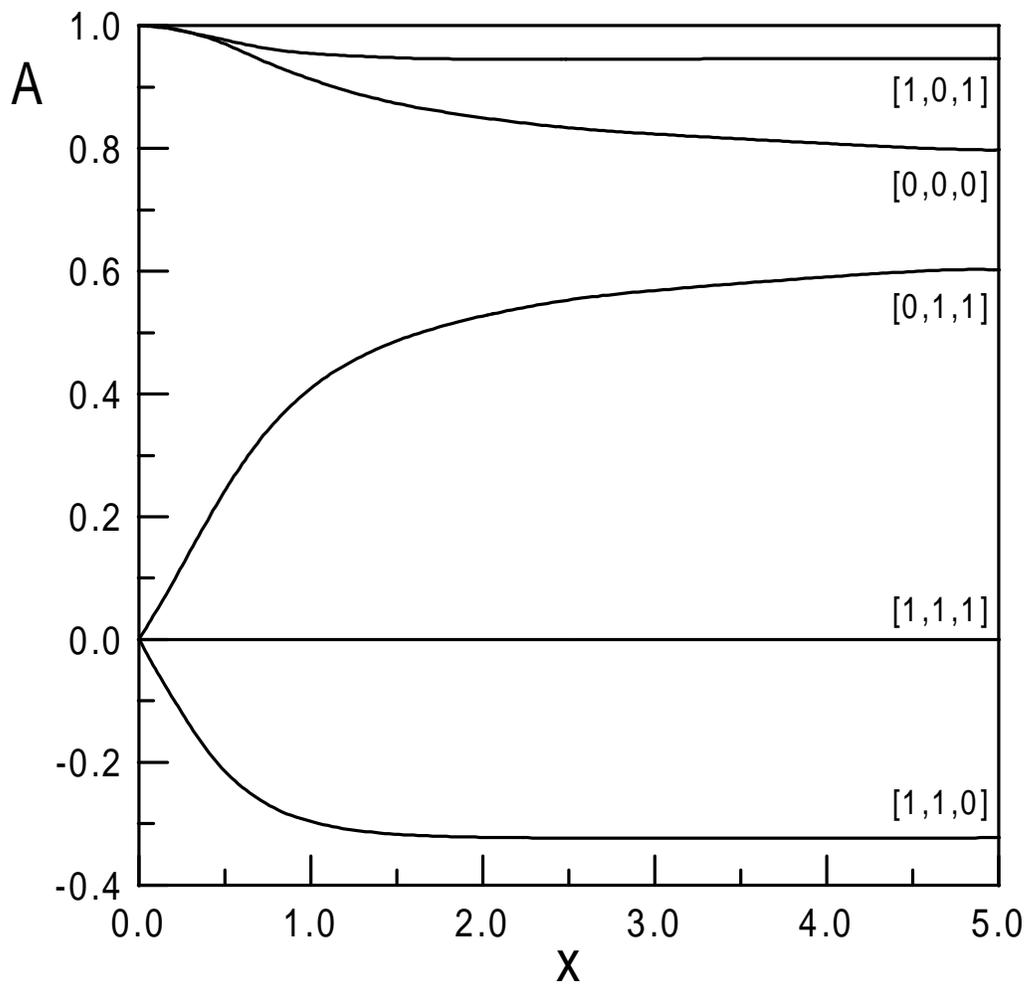
{\bf Fig.2} The MP parameters for the ladder $AA_{1/2}$.
In  figures 2 to 10 we adopt the notation 
$x = |J_\parallel/J_\perp |$. 
The curve $A_{J_1 J_2}^\lambda$
is labelled as $[J_1, J_2 , \lambda]$.

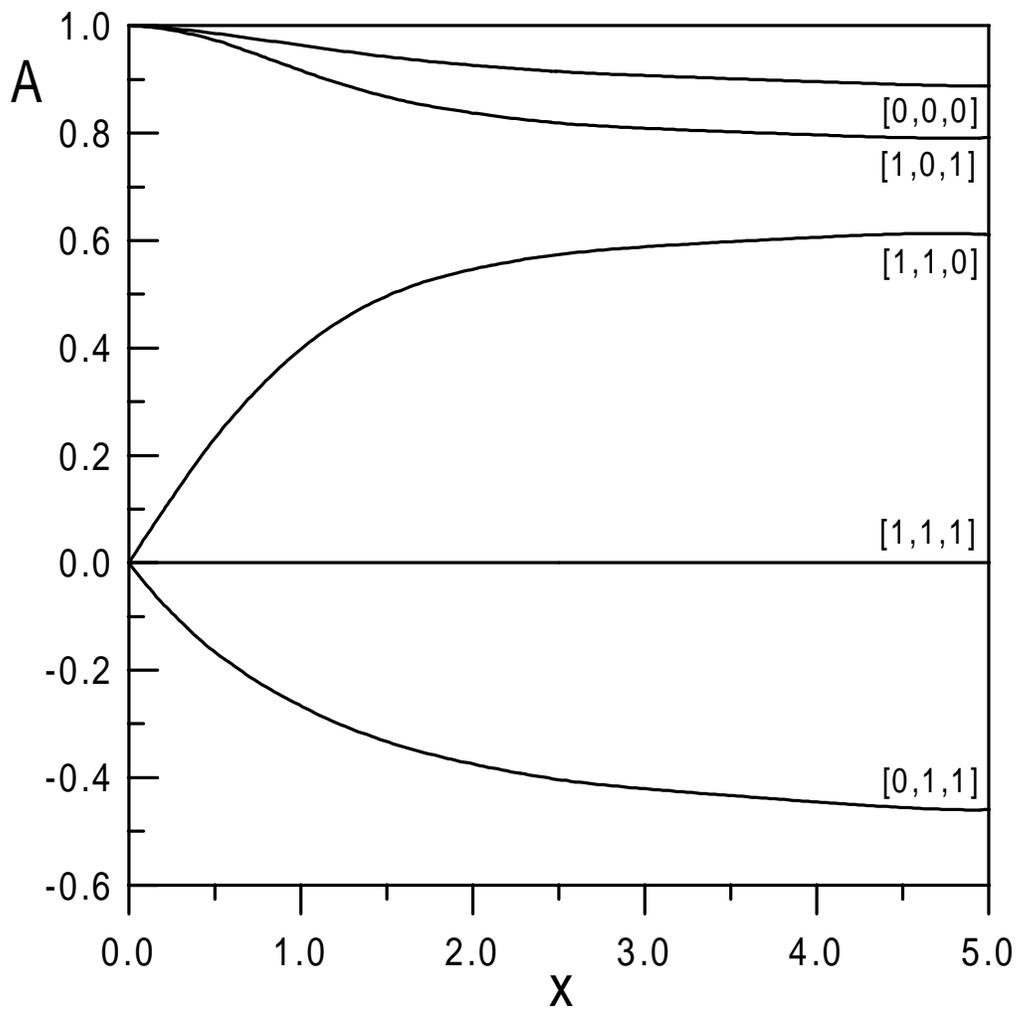
{\bf Fig.3} Same notations as in fig. 2 but for the $FA_{1/2}$ ladder.

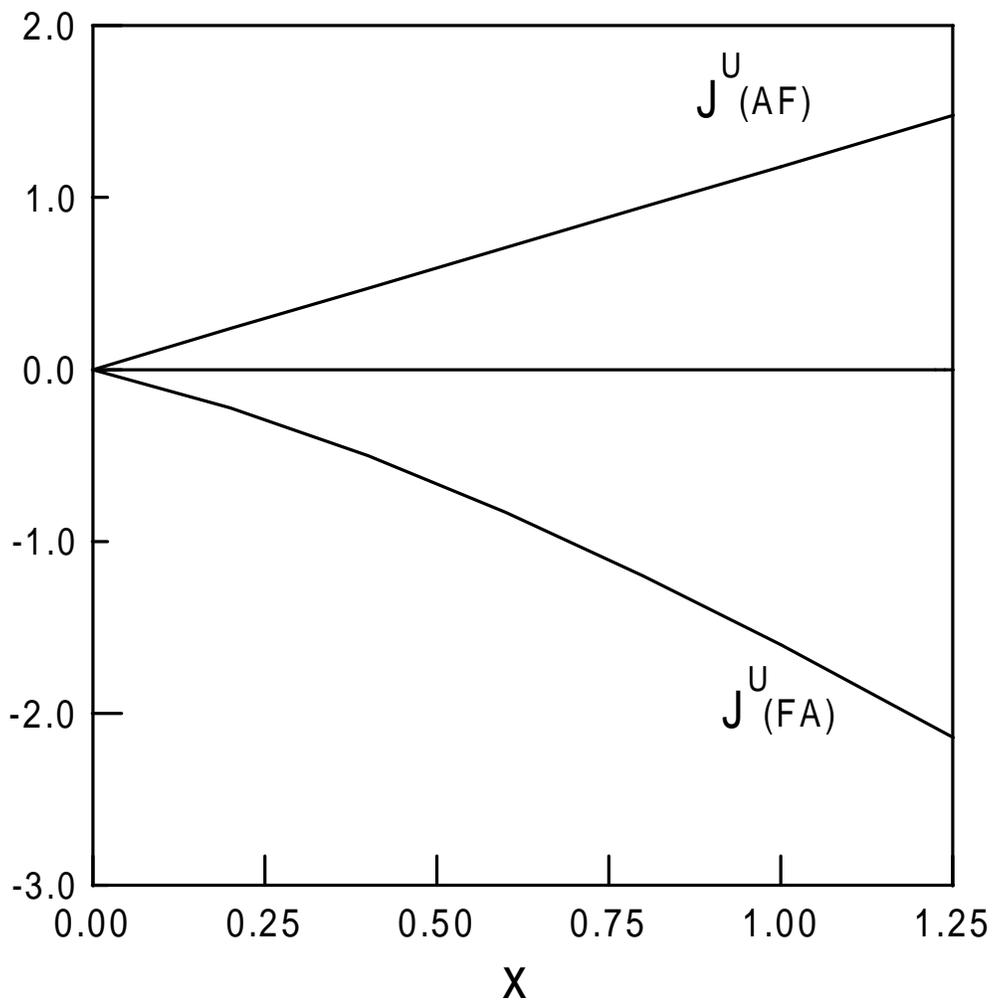
{\bf Fig.4} $J^U(AF) \equiv J^U_{\parallel}(AF)$ and
$J^U(FA) \equiv J^U_{\parallel}(FA)$.

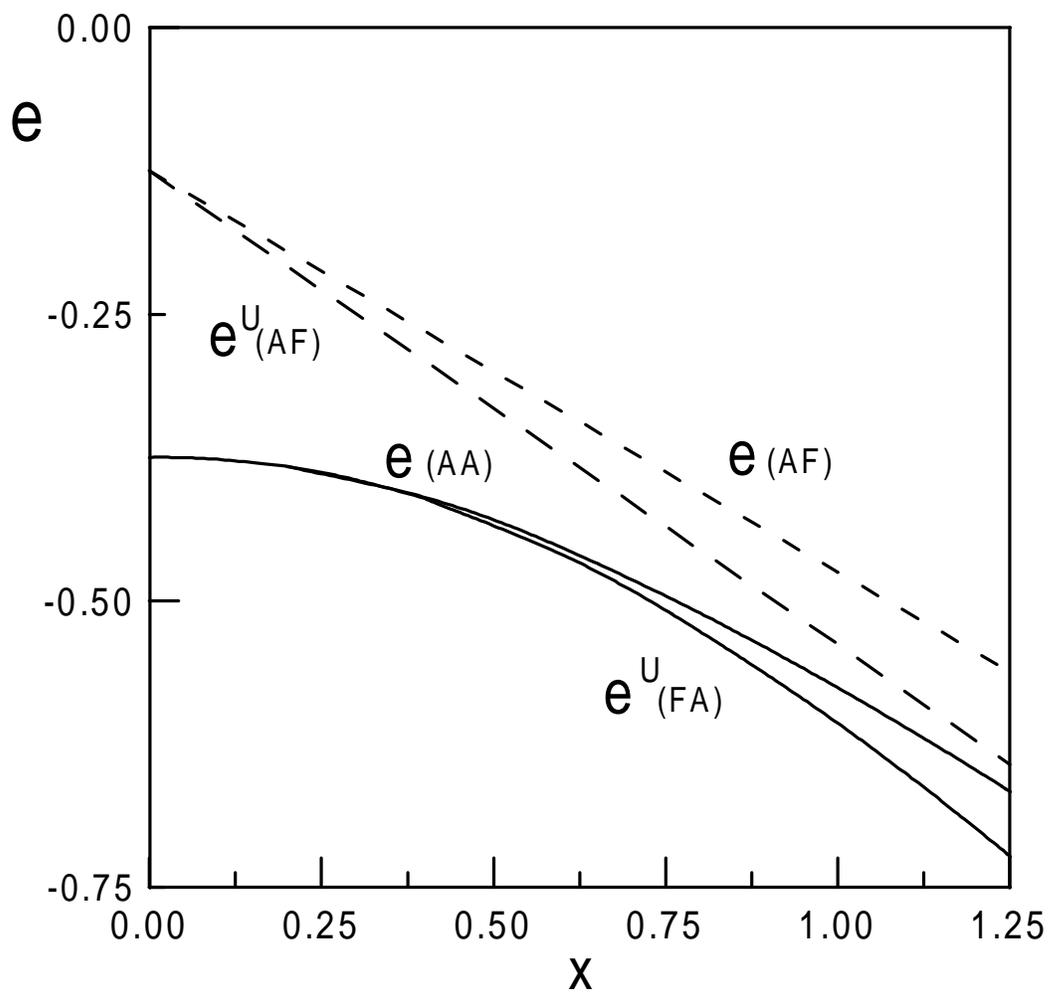
{\bf Fig.5} GS energy per site of the $AA$ and $AF$ ladders and 
their $U$ dual models.

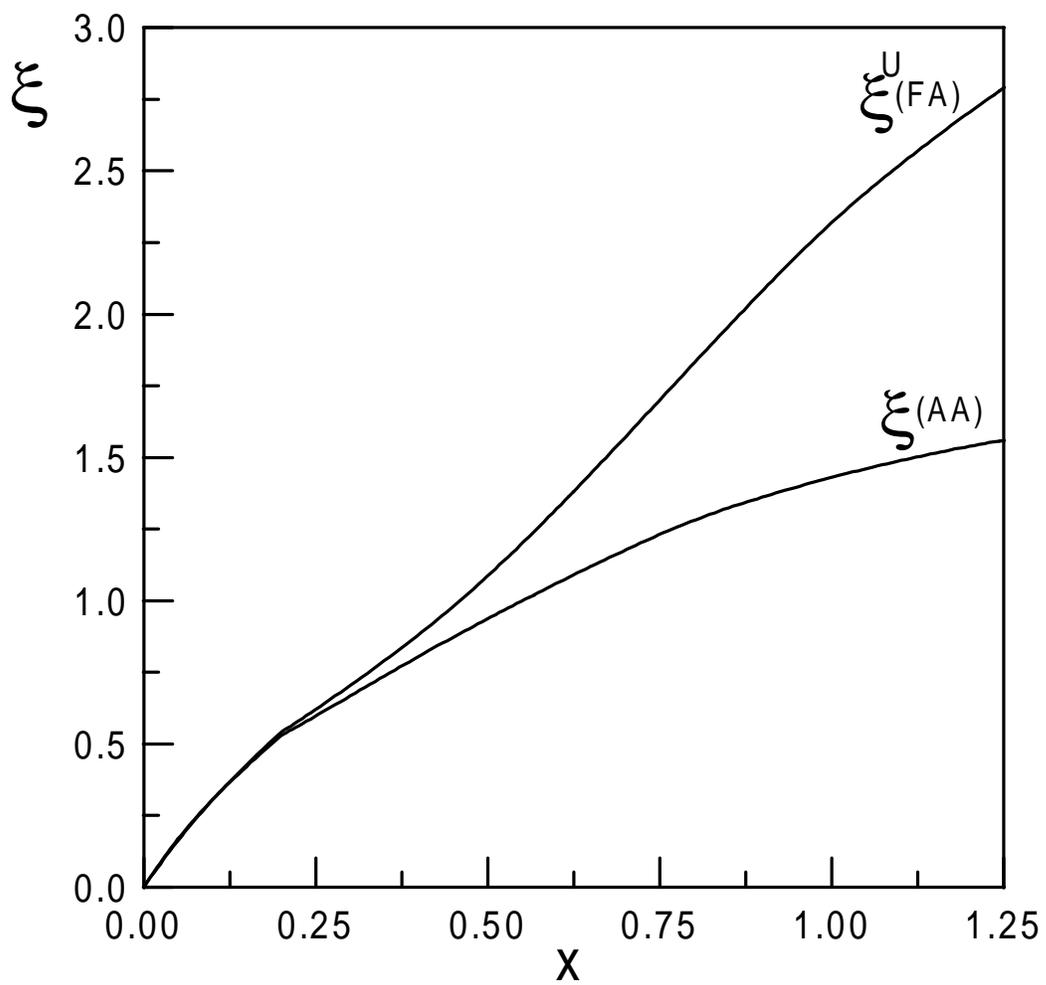
{\bf Fig.6} Spin-correlation  lengths 
of  the $AA$ ladder and its 
$U$ dual.

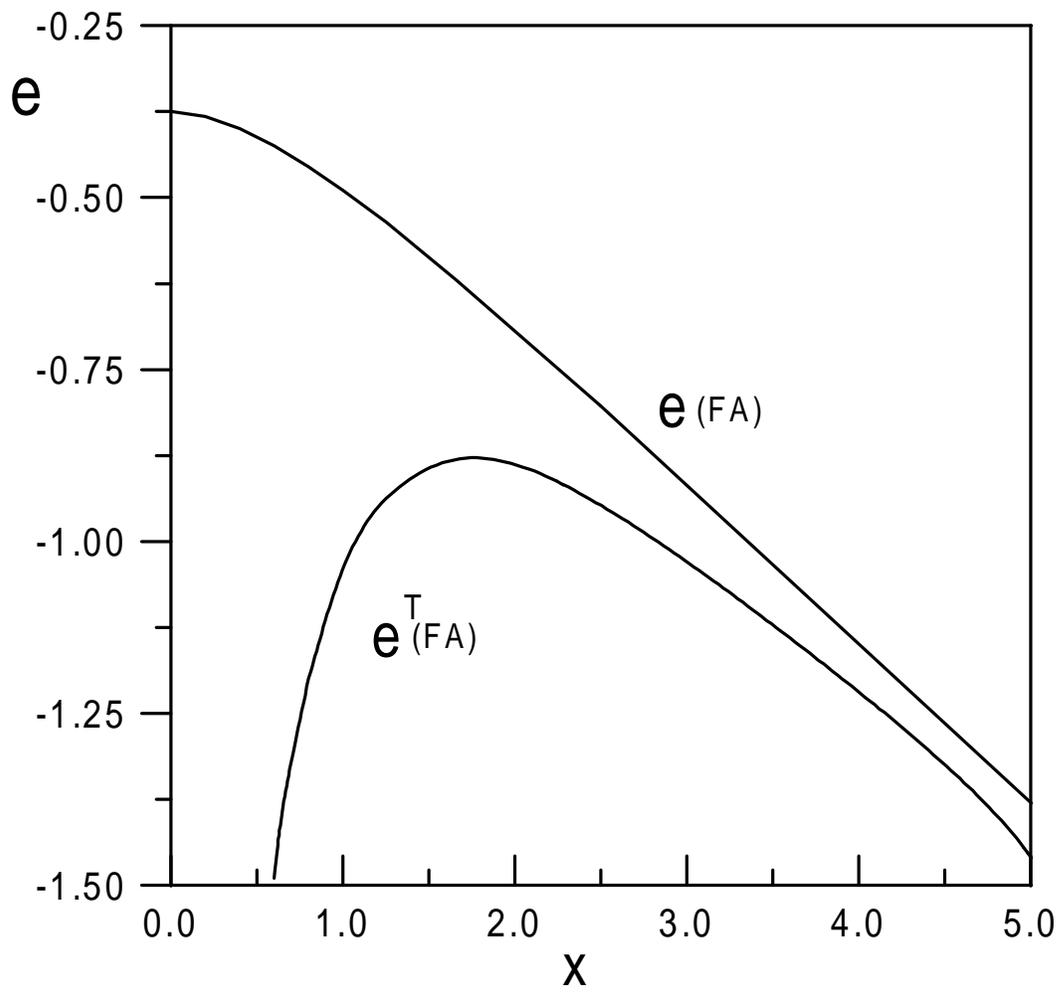
{\bf Fig.7} GS energy per site of the $AF$ ladder and
its  $T$ dual given by the $FA$ ladder.

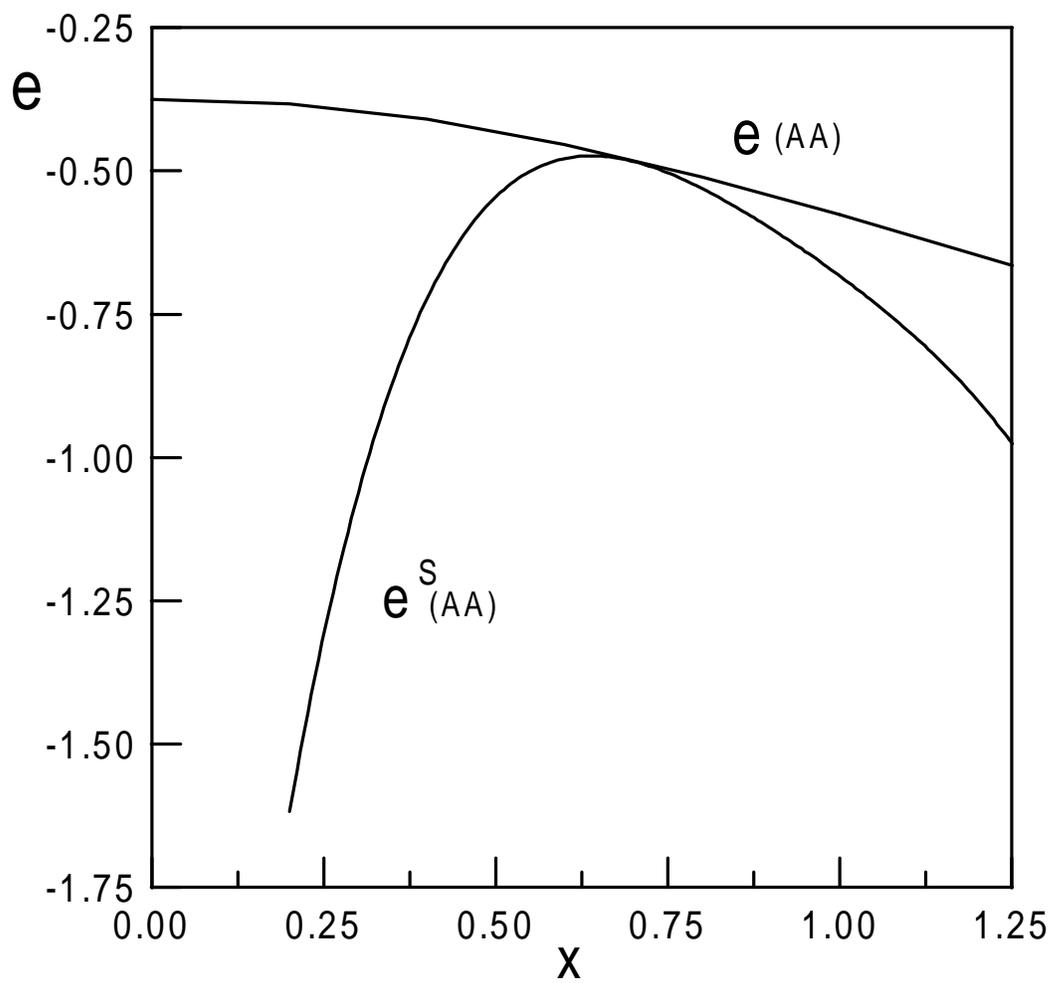
{\bf Fig.8} GS energy per site of the $AA$ ladder and 
its  $S$ dual.

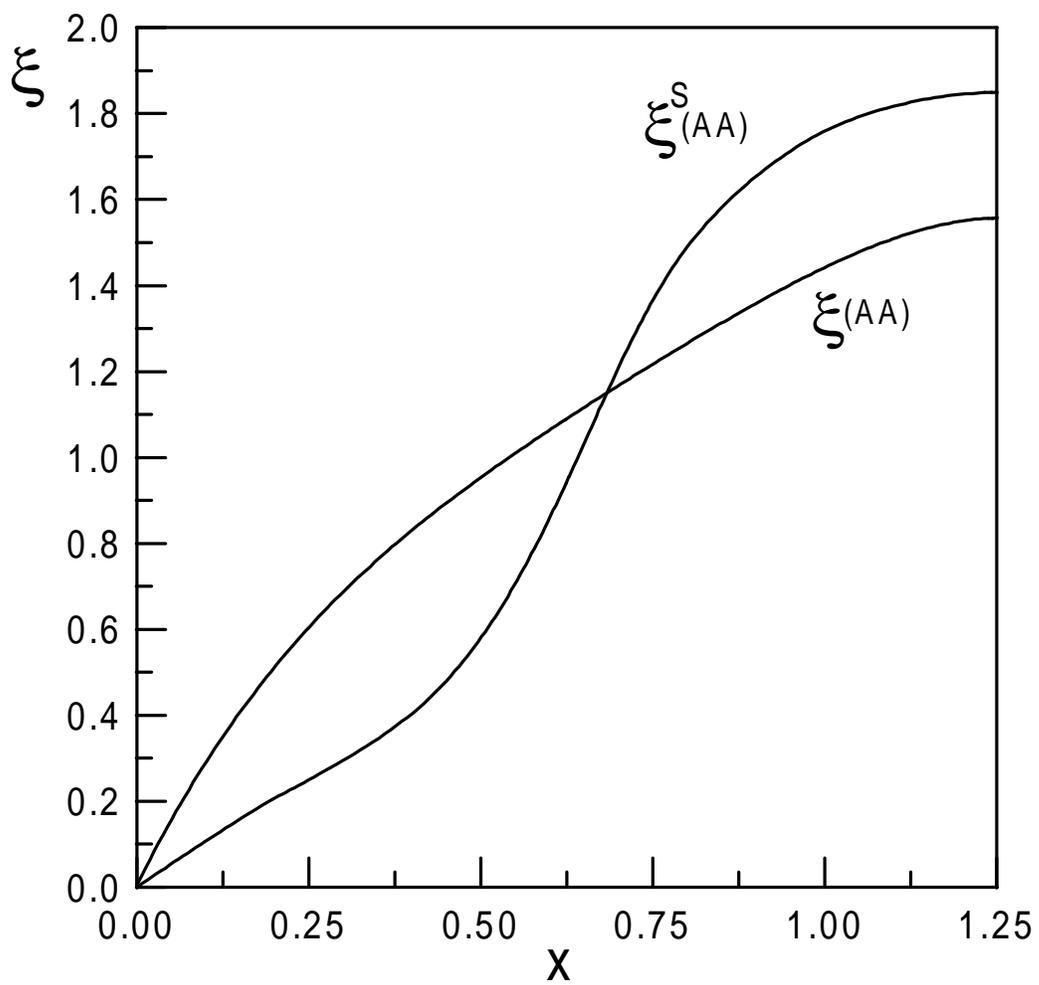
{\bf Fig.9} Spin-correlation length of the $AA$ ladder and 
its  $S$ dual.

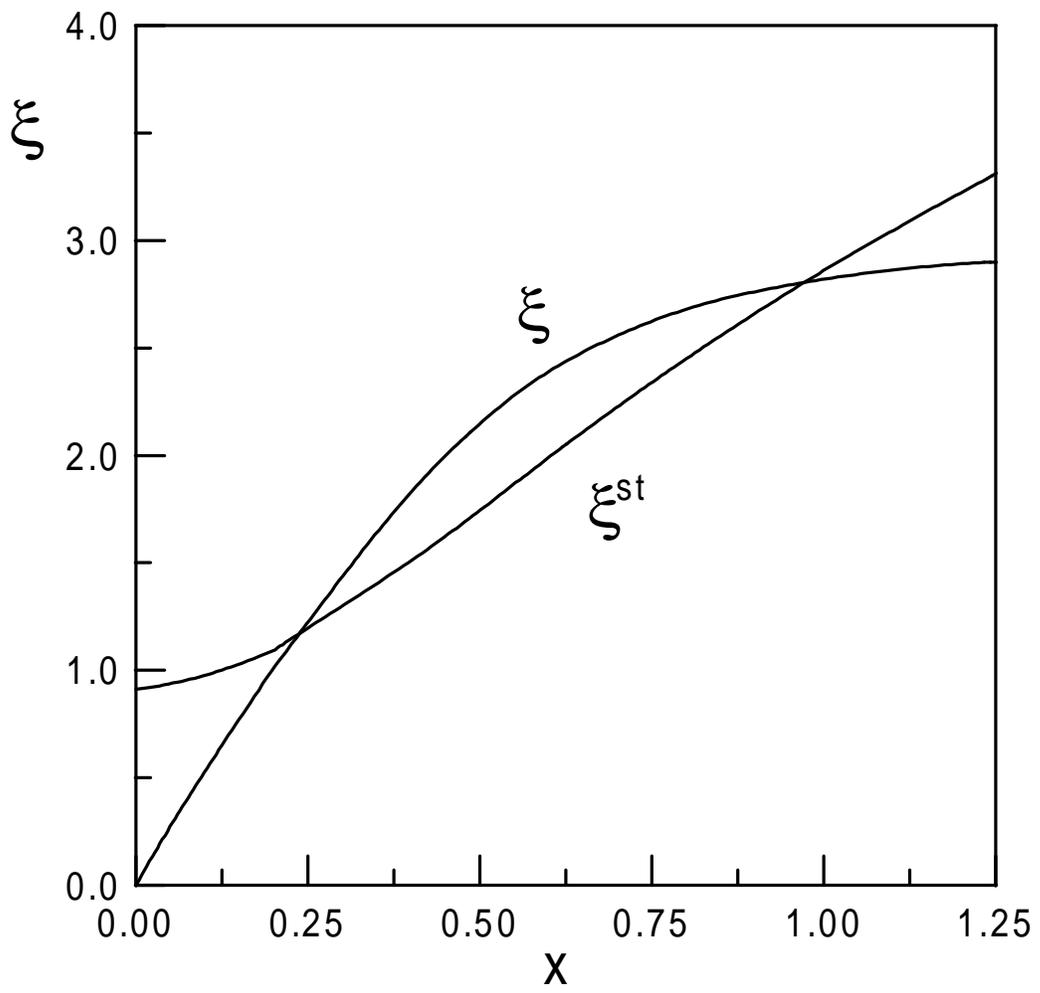
{\bf Fig.10} Plots of the spin-correlation length $\xi$
and the string correlation length $\xi^{\rm st}$ of the 
ladder $AA_1$.

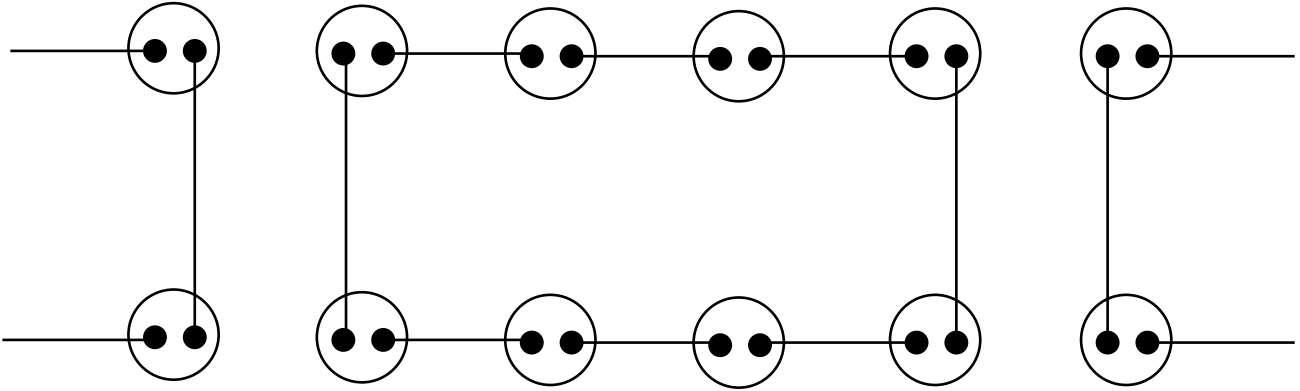
{\bf Fig.11} Pictorical representation of a possible AKLT state
of the spin 1 ladder.

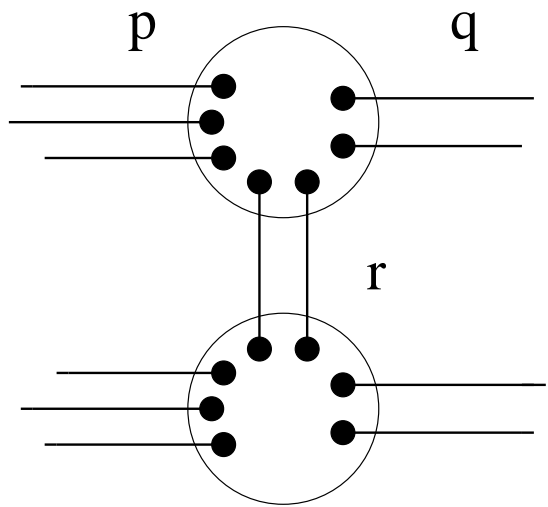
{\bf Fig.12} Graphical representation of a generic AKLT 
state of a ladder 
denoted as $(p,q,r)$. 
There are a total of $p+q+r$ dots inside every circle
representing a total spin $S= (p+q+r)/2$.

\end{document}